\documentclass{article}

\usepackage{PRIMEarxiv}

\usepackage[utf8]{inputenc} % allow utf-8 input
\usepackage[T1]{fontenc}    % use 8-bit T1 fonts
\PassOptionsToPackage{hyphens}{url}
\usepackage{hyperref}       % hyperlinks
\usepackage{booktabs}       % professional-quality tables
\usepackage{amsfonts}       % blackboard math symbols
\usepackage{nicefrac}       % compact symbols for 1/2, etc.
\usepackage{microtype}      % microtypography
\usepackage{lipsum}
\usepackage{fancyhdr}       % header
\usepackage{graphicx}       % graphics
\graphicspath{{media/}}     % organize your images and other figures under media/ folder

\usepackage{tabularx}
\usepackage{booktabs}
\usepackage{ragged2e}

\usepackage{enumitem}

\usepackage{verbatim}       % for comment
\usepackage{xcolor}         % for textcolor
\usepackage{amsmath}

\usepackage{float}

\usepackage{multirow}

\usepackage{soul}

\newcommand{\sunny}[1]{\textcolor{brown}{\textbf{Sunny:} #1}}
\newcommand{\yue}[1]{\textcolor{orange}{\textbf{Yue:} #1}}
\newcommand{\harsha}[1]{\textcolor{violet}{\textbf{Harsha:} #1}}

\title{Responsible AI Question Bank: A Comprehensive Tool for AI Risk Assessment}
\author{
  Sung Une Lee, Harsha Perera, Yue Liu, Boming Xia, Qinghua Lu, Liming Zhu, Olivier Salvado\thanks{Current affiliation: Queensland University of Technology, QLD, Australia. He conducted this work during his time at CSIRO.}, Jon Whittle \\
  Data61, CSIRO, Australia\\
}

\begin{document}
\maketitle

\begin{abstract}
The rapid growth of Artificial Intelligence (AI) has underscored the urgent need for responsible AI practices. Despite increasing interest, a comprehensive AI risk assessment toolkit remains lacking. This study introduces our Responsible AI (RAI) Question Bank, a comprehensive framework and tool designed to support diverse AI initiatives. By integrating AI ethics principles such as fairness, transparency, and accountability into a structured question format, the RAI Question Bank aids in identifying potential risks, aligning with emerging regulations like the EU AI Act, and enhancing overall AI governance. A key benefit of the RAI Question Bank is its systematic approach to linking lower-level risk questions to higher-level ones and related themes, preventing siloed assessments and ensuring a cohesive evaluation process. Case studies illustrate the practical application of the RAI Question Bank in assessing AI projects, from evaluating risk factors to informing decision-making processes. The study also demonstrates how the RAI Question Bank can be used to ensure compliance with standards, mitigate risks, and promote the development of trustworthy AI systems. This work advances RAI by providing organizations with a valuable tool to navigate the complexities of ethical AI development and deployment while ensuring comprehensive risk management.
\end{abstract}

% keywords can be removed

\section{Introduction} \label{sec:introduction}

%Importance of AI risk assessment
Since the emergence of ChatGPT and other large language models, Artificial Intelligence (AI) has seen a surge in popularity in recent years.
Fueled by remarkable advancements, companies across all industries, both global and regional, are rapidly adopting, using, and developing AI systems to enhance their businesses. 
While this rapid adoption has driven significant market growth in the AI industry and generated excitement for its potential, it has also raised concerns about the responsible development and application of AI, such as hallucination, generating harmful content, and user's overreliance \cite{OpenAI2023}.

A recent report found that while many companies view AI as a promising technology and actively pursue AI opportunities, only 10\% of those surveyed have publicly announced their Responsible AI (RAI) policies \cite{ESGAI}.
%\footnote{\url{https://www.csiro.au/-/media/D61/Responsible-AI/Alphinity/Responsible-AI-and-ESG.pdf}\label{ESG-AI report}}. 
This highlights that many companies are still concerned about their RAI maturity level.
Moreover, there have been numerous incidents involving AI \cite{aiincident}.
%\footnote{\url{https://incidentdatabase.ai/}}. 
AI incidents are not limited to specific industries like finance and healthcare. They have emerged across all sectors, raising various concerns in areas such as privacy, bias, and safety.

% Concept of RAI 
RAI is the practice of designing AI systems that deliver positive outcomes for individuals, groups, and society, while minimizing risks \cite{lu2022responsible}. 
AI development risks stem from misalignment with RAI principles such as fairness, privacy, explainability, and accountability. 
To mitigate these risks, it is crucial to ensure these principles are embedded throughout the AI development process \cite{xia2023towards}.

% background and concept of a question bank
There are a number of industrial RAI frameworks to support the implementation of RAI principles.
%Most of them are principle-driven and process-centred frameworks which provide a set of guidance for evaluating RAI practices and mitigating potential risks associated with AI systems.
In the US, the NIST AI Risk Management Framework \cite{nistairmf} %\footnote{\url{https://www.nist.gov/itl/ai-risk-management-framework}\label{NIST}} 
provides four functions such as MAP, GOVERN, MEASURE and MANAGE to manage AI risks. 
Recently, NIST has also released a new draft framework for Generative AI (GenAI) \cite{nistairmfgen}
%\footnote{\url{https://airc.nist.gov/docs/NIST.AI.600-1.GenAI-Profile.ipd.pdf}}
, which encompasses 12 potential risks of GenAI and provides comprehensive RAI practices to address those risks.
In Europe, the EU Assessment List for Trust AI Framework \cite{eualtai}
%\footnote{\url{https://futurium.ec.europa.eu/en/european-ai-alliance/pages/welcome-altai-portal}} 
has been used as AI ethics guidelines, which aligns with seven RAI principles (e.g., privacy, transparency, accountability) . 
In academia, researchers have increasingly focused on RAI practices and risk assessment for AI.

While existing frameworks provide valuable guidance, they often remain high-level and struggle to bridge the gap between policy goals and practical implementation in AI development.
%Moreover, there is a lack of a comprehensive and well-structured AI risk assessment frameworks. Especially, there is a lack of a RAI assessment framework that incorporates interconnected and layered risk questions and connectivity to other standards and frameworks 
Moreover, there is a lack of a comprehensive AI risk assessment framework that integrates interconnected, layered risk questions and aligns with other standards and frameworks \cite{xia2023towards}.
%\qinghua{mapping to stakeholders and system lifecycle stages are also our contribution?}
%\qinghua{how about system perspective}
%\sunny{added a sentence in the contricution section below}
Existing RAI frameworks often focus on specific aspects, failing to address the holistic nature of RAI principles. 
Liao et al. \cite{liao2020questioning, liao2021question} developed a question bank specifically focused on transparency and explainability, overlooking other crucial principles. 
Similarly, the US Department of Energy's AI Risk Management Playbook \cite{doeairmp}
%\footnote{\url{https://www.energy.gov/ai/doe-ai-risk-management-playbook-airmp}} 
provides extensive guidance on identifying AI risks and recommending mitigations, but it lacks integration with diverse standards, frameworks, and related resources.

% Concept of QB and objectives, contributions
To address current challenges, we present a novel RAI question bank developed based on a systematic literature review and in-depth survey on existing RAI frameworks.
The initial results in 2023 were suitable for general and broad applications \cite{lee2023qb4aira}.
A key advancement of this study is the integration of a systematic approach that links lower-level risk questions to higher-level ones and related themes to prevent siloed and fragmented assessments, ensuring a cohesive evaluation process across different levels of the organization. 
Additionally, it is to accommodate different levels of stakeholders (e.g., executives, senior managers, developers) and is mapped to the system life-cycle stages. This allows users to focus on their specific contexts and interests while maintaining a comprehensive system perspective, thus enhancing both the relevance and effectiveness of the risk assessments.
%Recent enhancements have significantly increased its flexibility, enabling support for a wide range of use case scenarios. These include compliance checks with various policies, such as the EU AI Act and ISO/IEC AI management standards, as well as risk assessments for AI agents and foundation models. 

% address Liming's comment
Although the RAI Question Bank relies primarily on yes/no questions, which risk fostering a checkbox mentality (i.e., the presence of low-quality artifacts or processes) if not carefully managed, we emphasize the importance of incorporating quality metrics and evidence-based responses. 
This means that the questions need to have quality aspects and answers are not just yes/no but require evidence, and lower-level answers collectively provide a crude metric for higher-level questions. 
Quality metrics and measurement are not the primary focus of this study.Yet, we have examined these considerations through a case study (Section \ref{sec:case study_2}) where we tested and refined the approach to ensure more meaningful and reliable assessments.

%\qinghua{does use case scenario here mean the mapping to standards? anything else?}
%\sunny{added AI agents}

\begin{comment} merged into the following
% Aim
This study presents a Responsible AI Question Bank, offering valuable insights and serving as a fundamental resource for AI risk assessment.
We also present two case studies to illustrate its practical application for real-world use and discuss how the question bank can be used to comply with regulations and build trust in the age of AI agents like Auto-GPT \cite{autogpt}.
\end{comment}

The paper is organized as follows.
Section \ref{sec:background} provides background information.
Section \ref{sec:methodology} describes the overall design and development approach of the RAI Question Bank.
We then present the Responsible AI Question Bank, offering valuable insights and serving as a fundamental resource for AI risk assessment (Section \ref{sec:questionbank}).
Section \ref{sec:case study} includes two case studies to illustrate its practical application for real-world use.
Section \ref{sec:discussion} covers the practical implications of this study, including how the question bank can be applied across various purposes and contexts, such as compliance checks and risk assessments of AI agents like Auto-GPT \cite{autogpt}.
We conclude this study in Section \ref{sec:conclusion}.

\section{Background} \label{sec:background}

The discourse on RAI has gained significant traction in both industry and academia, underscoring the critical need for effective AI risk management to foster RAI practices. Despite the proliferation of studies and frameworks on responsible and safe AI, many remain abstract, lacking concrete measures for risk assessment and management~\cite{xia2024towards}. Our previous mapping study, which served as a comprehensive literature review~\cite{xia2023towards}, systematically analyzed 16 AI risk assessment and management frameworks worldwide to gain insights into current practices in managing AI risks. This study revealed several key trends and areas for improvement in AI risk assessment practices globally, informing the design and development of our question bank.

\begin{itemize}
    \item \textbf{Increasing Global Concern}: The growing number of AI risk assessment frameworks worldwide indicates an increasing global concern about the risks associated with AI systems and a growing recognition of RAI approaches to assess and mitigate these risks.

    \item \textbf{Limited Scope and Stakeholder Consideration}: Many frameworks have a limited scope and fail to consider a comprehensive range of stakeholders and RAI principles, resulting in overlooked and unaddressed risks. %\boming{mention/add QB's stakeholder mapping and all 8 AU principles}
    %\sunny{added the following}
    To the best of our knowledge, there are no existing frameworks that cover all RAI areas, such as the eight AI ethics principles (Human/Societal/Environmental Wellbeing, Human-Centered Values, Fairness, Privacy/Security, Reliability/Safety, Transparency/Explainability, Contestability, Accountability). Furthermore, exisitng frameworks often do not encompass various stakeholders and system development phases necessary for a robust engineering approach. This study addresses these gaps by providing a comprehensive tool that integrates all these aspects, ensuring a thorough and inclusive AI risk assessment.

    \item \textbf{Lack of Context-Specific Guidance}: While frameworks are generally domain-agnostic and consider the entire lifecycle of AI systems, they lack clear guidance on adapting them to diverse contexts. This limitation hinders their effectiveness, as risks and mitigation measures can vary significantly depending on the specific context. %\boming{mention ESG-AI framework and other context-specific features/extensions?}
    %sunny(added the following}
    For instance, in the ESG (Environmental, Social, and Governance) context, AI can be used to mitigate ESG risks for a company (e.g., reduce carbon emissions) and can also generate positive impacts, including business, environmental, and social benefits. However, RAI risks and concerns often overlap with ESG risks and concerns. Therefore, RAI risk assessment can play a critical role in integrating ESG considerations with AI, ensuring that AI systems are not only effective but also aligned with broader ESG goals. We will show this case study in Section \ref{sec:case study}. 

    \item \textbf{Reliance on Subjective Evaluation}: Current frameworks primarily rely on subjective evaluations from assessors, which can lead to biased results. There is a need for the development and incorporation of more objective tools and techniques. 
    %\boming{can remove this one if QB doesn't address this either}
    %\sunny{added the following.}
    To support this, there should be clear instructions for users on how to evaluate the responses of examinees. In the second case study introduced in Section \ref{sec:case study}, we demonstrated how we have addressed this requirement in the RAI Question Bank.

    % \item \textbf{Insufficient Distinction of Risk Factors}: Frameworks do not clearly distinguish among different types of risk factors, such as hazard, exposure, vulnerability, and mitigation risk, which is crucial for effectively identifying and mitigating AI risks.

    %\item \textbf{Lack of Concrete Mitigation Solutions}: Many frameworks do not provide concrete mitigation solutions or lack a structured way to present them, making it challenging for organizations to effectively address identified risks. 
    %\boming{not sure if we will say / have said that QB is mapped to Patterns, which are mitigations}
    %\sunny{I blocked this}
\end{itemize}

%Building on these findings, we primarily focused on five specific frameworks that serve as the foundation for the initial version of the RAI Question Bank: 
Building on these findings, we have found five specific frameworks that can serve as the foundation for the development of the RAI Question Bank:

\begin{itemize}
    \item \textbf{NIST AI Risk Management Framework \cite{nistairmf}}: The NIST AI RMF offers a flexible and voluntary framework applicable across various sectors and use cases. It enhances AI system trustworthiness by focusing on four main functions: Govern, Map, Measure, and Manage. Developed through an inclusive and transparent process, the framework supports existing AI risk management initiatives and adapts to the evolving AI landscape, enabling organizations to implement it to different extents for societal benefit while minimizing risks. 

    \item \textbf{Microsoft Responsible AI Impact Assessment Template and Guide \cite{msraitemplate, msraiguide}}: Microsoft's resources support its Responsible AI Standard \cite{msraistandard}. They enable AI development teams to explore and understand the impacts of their AI systems on stakeholders, intended benefits, and potential harms from the earliest design stages.

    \item \textbf{EU Assessment List for Trustworthy Artificial Intelligence \cite{eualtai}}: ALTAI helps organizations self-assess the trustworthiness of their AI systems. This checklist translates key requirements from the Ethics Guidelines for Trustworthy AI \cite{eutrustworthyai} into actionable steps, ensuring alignment with principles to mitigate risks and promote responsible AI development and deployment.

    \item \textbf{Canada Algorithmic Impact Assessment \cite{canadaaia}}: This tool helps organizations evaluate the potential impacts of AI systems, especially in public services. It includes a detailed questionnaire to identify and mitigate risks associated with automated decision systems, focusing on bias, privacy, and transparency, and ensures compliance with ethical and legal standards.

    \item \textbf{Australia NSW AI Assurance Framework \cite{nswai}}: This framework ensures the safe, ethical, and responsible use of AI technologies in government services. It includes AI governance, risk management, and mitigation strategies, emphasizing stakeholder engagement and continuous monitoring to maintain compliance with ethical standards. It also addresses emerging AI risks, such as those posed by generative AI, ensuring AI projects meet community expectations and uphold public trust. 

\end{itemize}

% \textit{EU Trustworthy AI Assessment List, Canada Algorithmic Impact Assessment, Australia NSW AI Assurance Framework\, Microsoft (MS) Responsible AI Impact Assessment Guide, and the US NIST AI Risk Management Framework.}
% \boming{other frameworks and overall findings from the mapping study}

%The development of our question bank was further informed by the \textbf{EU AI Act}\footnote{\url{https://www.europarl.europa.eu/doceo/document/TA-9-2024-0138-FNL-COR01_EN.pdf}\label{EU AI ACT}} and \textbf{ISO\/IEC 42001:2023}\footnote{\url{https://www.iso.org/standard/81230.html}\label{ISO42001}}. 

AI risk assessments should account for the evolving landscape of AI regulations and standards such as \textit{the EU AI Act} \cite{euaiact} and \textit{ISO/IEC 42001:2023} \cite{iso}.

The EU AI Act establishes a comprehensive regulatory framework for AI within the European Union. It classifies AI systems based on their risk levels: unacceptable, high, limited, and minimal risk. Unacceptable risk AI applications are banned, while high-risk applications must meet stringent requirements for transparency, quality, and security, and undergo conformity assessments. Additionally, it establishes the European Artificial Intelligence Board to ensure compliance and promote cooperation among member states. 
ISO/IEC 42001:2023 specifies requirements for establishing, implementing, maintaining, and continually improving an AI management system within organizations. It is designed for entities providing or utilizing AI-based products or services, ensuring responsible development and use of AI. The standard focuses on quality management and reliability, guiding organizations in managing the issues and risks associated with AI technologies. This systematic approach facilitates consistency with other management system standards related to quality, safety, security, and privacy.

Advanced AI, including generative AI and foundation models (e.g., large language models (LLMs) such as GPT-4), presents significant challenges due to its complexity, scale, and potential for misuse~\cite{bengio2024managing}. These technologies can produce highly realistic text and images, raising concerns about misinformation, intellectual property violations, and other ethical dilemmas~\cite{shoaib2023deepfakes, govuk_ai_safety_interim_report, zhang2024privacy}. Effective risk management for these systems requires transparency, accountability, and rigorous evaluation processes~\cite{xia2024ai}. Comprehensive evaluations encompassing both AI and non-AI components at the model and system levels are essential for enhancing AI safety.
The added complexity introduced by environmental affordances, such as access to tools, user interactions, and safety guardrails, demands meticulous risk assessments for AI systems~\cite{anwar2024foundational}, including AI-based agents like LLM-agents~\cite{liu2024agent}, throughout their lifecycle—from inception to retirement.

Recent efforts to improve the safety of advanced AI include several notable works. Partnership on AI released guidance for the safe deployment of foundation models, targeting model developers \cite{paiguiance}. The AI Verify Foundation and Singapore's Infocomm Media Development Authority published the Model AI Governance Framework for Generative AI, providing high-level guidance on the design, development, and deployment of generative AI models \cite{aiverify}. On April 29, 2024, NIST released a draft of the AI RMF Generative AI Profile, guiding organizations in managing risks unique to generative AI \cite{nistairmfgen}. The EU AI Act also covers general-purpose AI like generative models, imposing specific transparency and evaluation requirements. The UK AI Safety Institute has published several resources, including Inspect \cite{aisiuk}, a platform for evaluating LLMs, and is actively developing infrastructure to assess the safety and societal impacts of advanced AI. Recently, the AU NSW Government updated its Artificial Intelligence Assessment Framework \cite{nswai}. The revised framework, formerly known as the NSW Artificial Intelligence Assurance Framework, now categorizes generative AI-based solutions as "elevated risk," requiring additional and stricter assessments.

% as specific types of AI systems, AI-based agents (e.g., LLM-agents~\cite{liu2024agent}) utilizing these advanced models require meticulous risk assessments focused on their decision-making processes and user interactions. Ensuring adherence to ethical guidelines involves scenario analysis, user feedback loops, and the establishment of clear ethical boundaries to prevent harmful outcomes

In summary, addressing the risks associated with AI, especially advanced AI, requires a multifaceted approach that integrates technical, ethical, and governance strategies to ensure responsible and beneficial deployment. Our paper aims to address this gap by contributing a comprehensive question bank for AI risk assessment to enhance RAI and AI safety. This question bank offers a structured approach for identifying and assessing %, and \boming{mitigating (not sure)} 
AI risks, providing stakeholders with practical tools to operationalize RAI. By focusing on concrete risk management measures, we aim to bridge the gap between theoretical frameworks and practical implementation, thereby enhancing the overall coherence and applicability of responsible AI practices.

% \begin{itemize}
%     \item 5 frameworks 
%     \item EU AI Act
%     \item ISO standards
%     \item AI safety standard (draft..... exclude?)
%     \item Other research works (from mapping study)
%     \item Comparison with other QB/framework ?????  our coverage/features....
% \end{itemize}

% \sunny{Boming and Yue, I leave this section for you. You can freely orgnize this section.}

\section {Methodology} \label{sec:methodology}
% \begin{itemize}
%     \item Mapping study methodology 
%     \item additional (identifying requirements from regulations/standards)
%     \item case study 
% \end{itemize}

\begin{figure*}[htb]
    \centering
    \includegraphics[width=\textwidth]{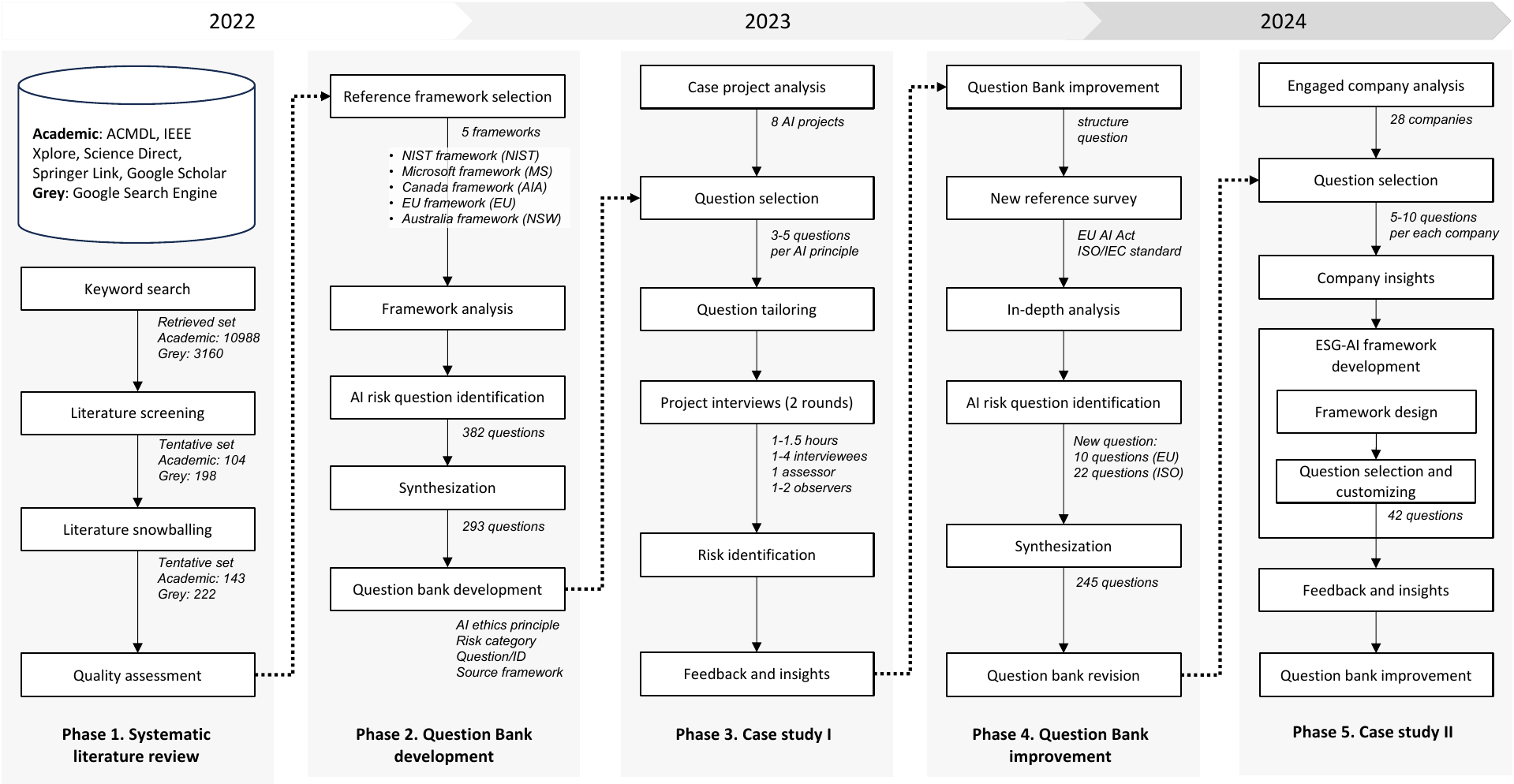}
    \caption{Research methodology overview.}
    \label{fig:methodology}
\end{figure*}

This study was conducted in five dedicated phases from 2022 to 2024 (Figure~\ref{fig:methodology}). 
First, a systematic mapping study was performed to understand the state-of-the-art in AI risk assessment and select reference frameworks \cite{xia2023towards}.
We selected five frameworks from the results, scrutinized, and extracted all included AI risk questions. The collected questions were synthesized and used to develop a comprehensive and holistic question bank for AI risk assessment~\cite{lee2023qb4aira}.
We then conducted a case study with eight AI development projects. Some selected questions were tailored for each project and used for interviews with team members. We gained insights and feedback from the participating project teams. Based on the feedback, we improved the initial version of the question bank.
As new AI regulations and standards are widely adopted, we incorporated the EU AI Act and ISO/IEC AI management standard to support organizations in achieving compliance.
Finally, we evaluated the feasibility and applicability of the question bank through a case study of the ESG-AI framework development project.

\subsection{Systematic Literature Review}
The systematic literature review was conducted adhering to guidelines for systematic multivocal literature review~\cite{GAROUSI2019101, KITCHENHAM20097} and mapping study~\cite{PETERSEN20151}. In particular, we aim to comprehend the extant frameworks for AI risk assessment; e.g., what the characteristics (e.g., demographics, stakeholders, scope) of existing AI risk assessment frameworks are and how the AI risks are assessed (i.e., input/output, assessment process).

\begin{comment}
\begin{itemize}
    % \item RQ1: Who have published RAI risk assessment frameworks?
    \item RQ1: What are the characteristics (e.g., demographics, stakeholders, scope) of existing AI risk assessment frameworks?
    % \begin{itemize}
    %     \item RQ1.1 What are the demographics of the frameworks?
    %     \item RQ1.2 What RAI principles are addressed in the frameworks?
    %     \item RQ1.3 Who are the stakeholders?
    %     % \begin{itemize}
    %     %     \item RQ1.3.1 Who conducts the assessment?
    %     %     \item RQ1.3.2 Whose activities are assessed?
    %     % \end{itemize}
    %     \item RQ1.4 What is the scope of the frameworks?
    %     % \begin{itemize}
    %     %     \item RQ1.4.1 Which development stages are covered by the frameworks?
    %     %     \item RQ1.4.2 Where can the frameworks be applied?
    %     %     \item RQ1.4.3 Which domains/sectors are the frameworks designed for?
    %     % \end{itemize}
    % \end{itemize}
    \item RQ2: How are the AI risks assessed, i.e., input/output, assessment process?
    % \begin{itemize}
    %     \item RQ2.1: What are the inputs?
    %     \item RQ2.2: What is the assessment process?
    %     \item RQ2.3: What are the outputs?
    % \end{itemize}
\end{itemize}
\yue{omit the sub-sub-questions..}
\sunny{I think mapping study part doesn't need to be detailed too much.}
\yue{Sure, we can shorten a little bit.}
\end{comment}

We collected the literature from the following sources: ACM Digital Library, IEEEXplore, ScienceDirect, SpringerLink and Google Scholar for academic paper, and Google Search Engine for grey literature. We selected the academic data sources as they are all recognized as the most representative digital libraries for Software Engineering research~\cite{KITCHENHAM20097}. We performed the search and the selected keywords and supportive terms are as follows: i) \textit{AI}: artificial intelligence, machine learning, ML; ii) \textit{risk}: impact; iii) \textit{assessment}: assess, assessing, evaluate, evaluation, evaluating, measure, measurement, measuring, mitigate, mitigation, mitigating, manage, managing, management. 
% The original search resulted in 10,988 academic papers and 3,160 grey literature. We then conducted a full-text screening and selected a tentative set of primary studies. Specifically, the inclusion and exclusion criteria were formulated as follows:

% \textbf{Inclusion criteria:}

% \begin{itemize}
%     \item Literature that discusses risk assessment on AI with relative concrete solutions;

%     \item Literature that is in the form of a published scientific paper or industry publication;

%     \item Frameworks are currently being used in practice (e.g., governmental/industrial/international organisations consulting extensively with practitioners and extracting proven uses).
% \end{itemize}

% \textbf{Exclusion criteria:}

% \begin{itemize}
%     \item Literature that only discusses the AI bias or issues, without assessing the risk of such issues;

%     \item Literature that generally discusses AI risks without more detailed solutions;

%     \item Literature that discusses leveraging AI for risk management in different fields;

%     \item Literature that only discusses risk assessment but not about AI;

%     \item Literature which is not written in English;

%     \item Conference version of a study that has an extended journal version;

%     \item Survey and review papers.

%     \item PhD/Master's dissertations, tutorials, editorials, books.
% \end{itemize}

We carried out a forward and backward snowballing process to include the related studies that might be missed in the initial search. Hereby, backward and forward snowballing means inspecting the references and citations of literature to search for related studies~\cite{snowballing}. In this process, 63 studies were included. Afterwards, an assessment was conducted to evaluate their quality and finalize the inclusion eligibility. Four quality criteria were developed for academic papers, and six for grey literature. Each criterion can be answered by one of three scores: $1$ (yes), $0.5$ (partially), and $0$ (no). We excluded academic papers scored below $2$ and grey literature scored below $3$. Finally, 16 grey literature were selected for data extraction and synthesis. 
%Our findings and insights are discussed in the previous section.
% The quality criteria are as follows:

% \textbf{For academic papers}:

% \begin{itemize}
%     \item Are the aims of the research clearly stated?

%     \item Is the research design appropriate and justified?

%     \item Are the findings and contributions of the study clearly stated and supported?
    
%     \item Are the limitations of the study or possible future work adequately described?
% \end{itemize}

% \textbf{For grey literature}:

% \begin{itemize}
%     \item Is the publishing source (e.g., organization, author(s)) reputable?

%     \item Does the source have a clearly stated aim?

%     \item Does the source have a clearly stated methodology?

%     \item Is the source supported by authoritative, documented references?
    
%     \item Does the framework cover specific/concrete Al risk assessment solutions?
    
%     \item Outlet type:

%     \begin{itemize}
%         \item 1st tier (measure = 1): High outlet control/High credibility: Books, magazines, theses government reports, white papers/reports.

%         \item 2nd tier (measure = 0.5): Moderate outlet control/Moderate credibility: Annual reports, news articles, videos, Wiki articles.

%         \item 3rd tier (measure = 0): Low outlet control/Low credibility: Blog posts, presentations, tweets
%     \end{itemize}
% \end{itemize}

% The major findings of this mapping study are as follows:

% \yue{@Boming please summarise the takeaways/insight of mapping study here}
% \boming{will add in background}

\subsection{Question Bank Development}

Based on the previous phase, five frameworks out of the 16 frameworks was chosen for further analysis and synthesis. 
The frameworks were selected according to i) global recognition, ii) inclusion of risk assessment questions, and, iii) representation of different regions and industry leaders. The selected frameworks include: US NIST AI Risk Management Framework \cite{nistairmf}, EU Trustworthy AI Assessment List \cite{eutrustworthyai}, Canada Algorithmic Impact Assessment \cite{canadaaia}, Australia NSW AI Assurance Framework \cite{nswai}, and Microsoft Responsible AI Impact Assessment Guide \cite{msraiguide}. By investigating these frameworks, we can comprehend and synthesize risk assessment solutions throughout various geopolitical contexts.

%\yue{please check the whether the link is for the included EU Trustworthy AI Assessment List. If yes I will add it to footnote. https://op.europa.eu/en/publication-detail/-/publication/73552fcd-f7c2-11ea-991b-01aa75ed71a1  }

We began by extracting 382 questions for risk assessment, which were subsequently refined through an iterative process to consolidate questions on similar topics. This refinement resulted in a condensed question bank of 293 questions. We utilized Australia's AI ethics principles \cite{auaiethicsprinciple} to categorize these questions. 
We structured the hierarchy of our question bank using a decision tree~\cite{song2015decision}, organizing questions based on their levels (level 1-3) and sequences to ensure a coherent and comprehensive framework.
To address potential gaps in technical/practical detail at lower level questions, we recognized that some lower-level (level 2 and 3) questions can be inferred from broader high-level frameworks. We made inferences to generate or extract these detailed questions where necessary. 
Concept mapping~\cite{TROCHIM19891} was employed to identify common RAI themes, allowing us to further analyze and group the questions.

%\yue{Rewrite based on the IEEE Software QB paper. Are there more details that need to specify?}

%\yue{the previous paper mentioned "stage" was added after case study, this can be stated in Section 3.4}

%\yue{Please check the highlighted number, I see Sunny mentions "update" in Section 4.}

\subsection{Case Study} 

The evaluation of our proposed question bank was carried out via two phases of case study (Case study I and II- Phase 3 and 5 in Figure \ref{fig:methodology}). 

The initial case study (Phase 3) with eight AI projects was conducted in 2023. 
We analyzed these projects to understand their contexts and selected 3 to 5 questions from our question bank for each AI ethics principle to prepare interviews. 
In parallel, we prepared a dedicated risk register template, including risk ID, category, title, description, causes, and selected interview questions derived from the question bank. 
For each project, we scheduled a 1.5-hour interview. 
We synthesized and analyzed the interview notes to identify and assess the AI ethics risks the project teams might encounter. 
We also asked the interviewees for feedback on the question bank for further refinement. 
We conducted two rounds of interviews in 2023 and consolidated the feedback to improve our question bank.
The second case study (Phase 5) was conducted with the ESG-AI investor framework development project, following the question bank improvement (Phase 3), based on the initial case study results and changes in the external AI regulation landscape. 
This case study engaged 28 companies using and developing AI. Considering the companies' characteristics and contexts, we selected and customized 5-10 questions for the interviews. 
In the ESG-AI framework development, 42 questions from the RAI Question Bank were selected to contribute to a deep dive assessment of companies. During the project, we received user feedback, improved the RAI Question Bank accordingly, and discussed the next steps.

\subsection{Question Bank Improvement}

Based on the results from the case studies, we enhanced the question bank by reviewing and refining its structure and adding new features. For example, we mapped the questions to development phases and incorporated principle-level questions for the eight AI ethics principles. Additionally, we revised the risk categories and questions to improve clarity and usability. Consequently, the RAI Question Bank now includes 26 main risk categories and 245 questions.

Another key improvement is the incorporation of AI regulations and standards. We conducted an in-depth analysis of the EU AI Act and ISO/IEC AI management standard to include relevant questions in the RAI Question Bank. The EU AI Act was chosen as it is the first comprehensive AI law worldwide, and the ISO 42001:2023 standard was included due to its widespread industry acceptance.
One researcher initially scrutinized these regulations and standards, extracting and enumerating requirements to be used for potential risk questions. Three researchers then reviewed and validated the results. As a result, we identified 25 requirements from the EU AI Act, ultimately adding 10 new questions to the RAI Question Bank after removing overlaps with exisiting questions in the RAI Question Bank. 
This included specified requirements and questions regarding General-Purpose AI Models (e.g., foundation models), which were previously not deeply analyzed in our question bank. Similarly, we identified 30 requirements from the ISO standard and added 22 questions to the RAI Question Bank.

This approach had two main objectives: i) to support the potential use of the RAI Question Bank for compliance purposes, and ii) to facilitate the operationalization of high-level requirements in regulations and standards.
We validated that the RAI Question Bank conforms to mainstream regulations and industry standards and complemented the question bank with more detailed RAI questions. 
Consequently, the RAI Question Bank can be used to interpret regulations and standards into low-level assessment checklists, enabling stakeholders to assess their compliance with corresponding requirements. In Section \ref{sec:complaince}, we detail how users can perform compliance checking with the RAI Question Bank.

\section{Responsible AI Question Bank} \label{sec:questionbank}

\begin{comment}
\begin{itemize}
    \item Overview of the architecture
    \item Key features (from the previous paper)
    \item Mapping with regulations (from NAIC project)
\end{itemize}
\end{comment}

\subsection{Overview}

This section presents the RAI Question Bank which is a set of risk questions designed to support the ethical development of AI systems. 
It addresses the needs of various users, including C-level executives, senior managers, developers, and others.

The purpose of the RAI Question Bank is to address the unique challenges and considerations associated with AI risk assessment while providing a comprehensive and effective tool for this purpose. 
The necessity for such a question bank is evident, given the complex and multifaceted nature of AI risks that can arise from various sources, including biased data, algorithmic errors, unintended consequences, and more. 
To effectively manage the complex and multifaceted nature of AI risks, organizations require a structured and systematic approach that encompasses all relevant areas of risk management. 

\begin{figure*} [htb]
    \centering
    \includegraphics[width=\textwidth]{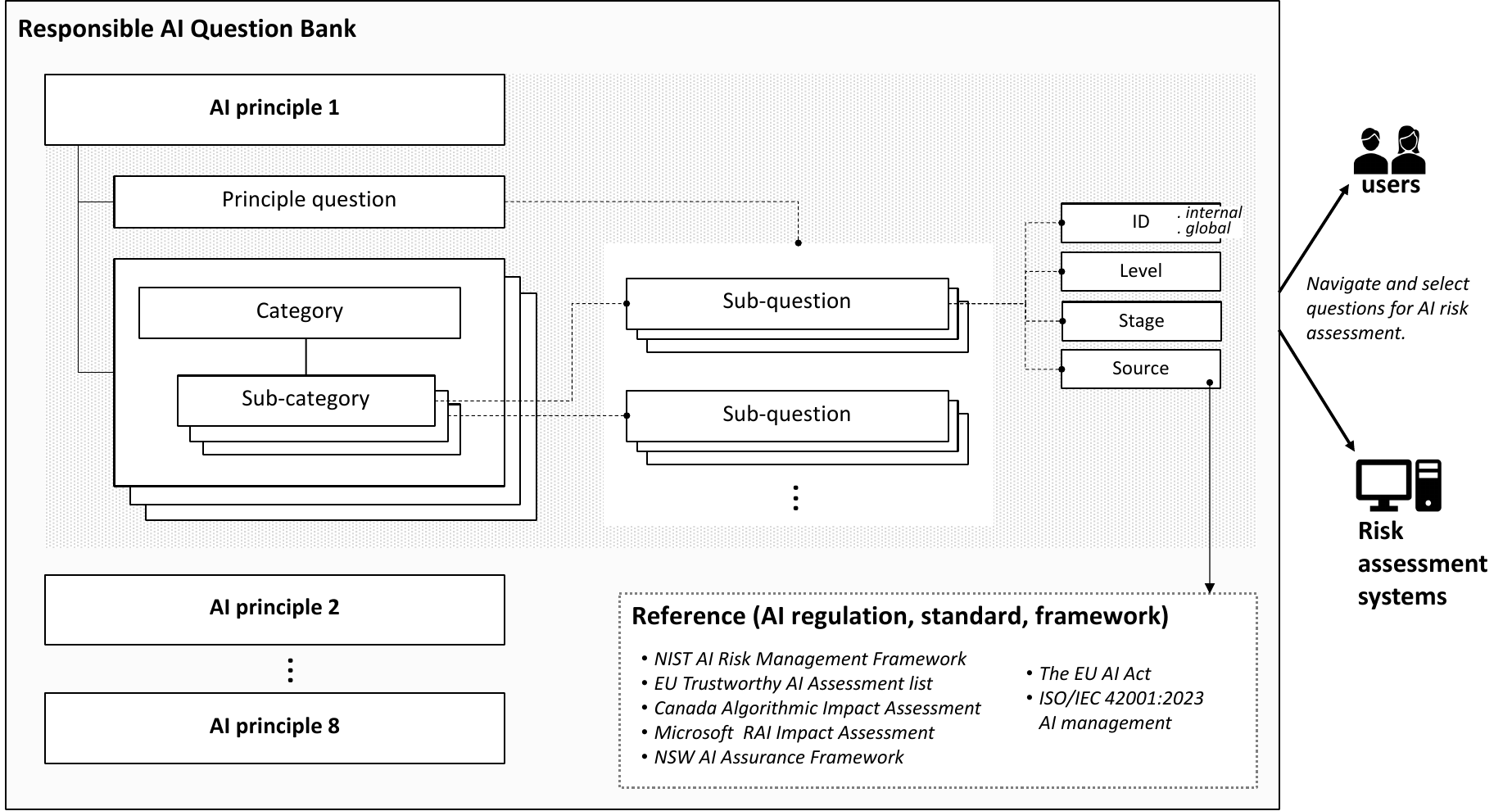}
    \caption{The architecture of the question bank.}
    \label{fig:QB}
\end{figure*}

Figure \ref{fig:QB} presents the overall architecture of the RAI Question Bank.
It is based on AI ethics principles which serve as an overarching category for the question bank.
Specifically, we have selected \textit{Australia's AI ethics principles} \cite{auaiethicsprinciple} to comprehensively cover responsible AI concerns.
The principles closely align with similar principles \cite{jobin2019global, fjeld2020principled} from around the world.
The principles comprise eight principles such as \textit{Human, Societal and Environmental Wellbeing}, \textit{Human-centred Values}, \textit{Fairness}, \textit{Privacy and Security}, \textit{Reliability and Safety}, \textit{Transparency and Explainability}, \textit{Contestability}, and \textit{Accountability}.
Each principle has one principle question, which can be used to get high-level insights and understanding of RAI practices and risks associated with the principle.
\textit{The principle question} includes a set of sub-questions which are distributed across risk categories within the principle.

\textit{Categories} were identified by conducting concept mapping to effectively comprehend relevant RAI concepts \cite{trochim1989introduction}. The concept mapping approach was used to analyze and group related risk questions, thereby identifying RAI themes. Through this process, we identified gaps and areas for improvement (e.g., missing information, duplicated categories, structural issues), clarified the relationships between different themes, and provided a comprehensive overview of the key areas of concern for AI systems. 
Each category is exclusively associated with one principle.
As shown in Figure \ref{fig:QB}, principle 1 has been further divided into multiple categories, each representing a specific aspect of the principle. 
The categories act as common RAI themes that are clear, specific, and relevant to the AI ethics principle they represent. Additionally, a category  contains one or more sub-categories, providing even more granularity and specificity to the questions and facilitating a better understanding of the different dimensions of each category.

\begin{figure*} [htb]
    \centering
    \includegraphics[width=\textwidth]{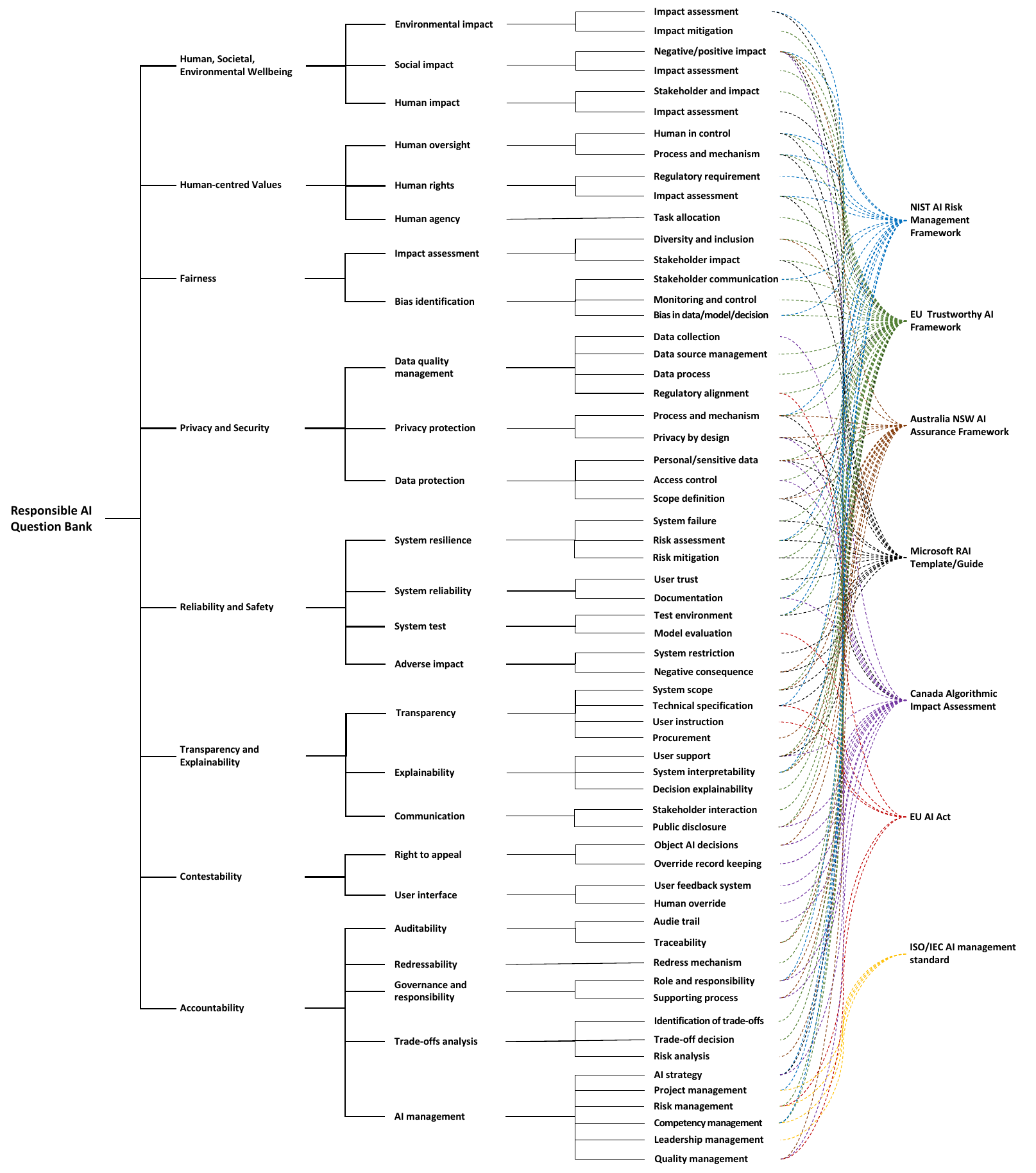}
    \caption{The RAI Question Bank categorization, comprising 26 categories and 65 sub-categories, identified from seven AI frameworks.}
    \label{fig:concept map}
\end{figure*}

Figure \ref{fig:concept map} represents the hierarchical organization of RAI risk categories including 26 categories and 65 sub-categories and their distribution across eight AI principles. 
The figure aims to visualize the structure of the RAI Question Bank in a clear and logical manner, providing a comprehensive overview of the taxonomy of RAI risk assessment areas derived from seven existing AI frameworks.
This hierarchical representation not only facilitates the understanding of the relationships between AI ethics principles, categories, and sub-categories but also enables a systematic exploration of the key areas within RAI risk assessment. By organizing these elements visually, the figure serves as a guide for navigating complex ethical considerations, making it easier to identify how different risk areas are interconnected and aligned with existing AI frameworks.
Additionally, it includes links between each sub-category and its corresponding source frameworks, illustrating how the seven existing AI frameworks have contributed to and informed the development of this taxonomy. 
As depicted in the figure, while most frameworks provide comprehensive coverage of the categories, the ISO standard emphasizes \textit{Accountability}, focusing primarily on management aspects. Similarly, the EU AI Act places greater emphasis on \textit{Accountability} and \textit{Transparency}.
This visual representation allows for a clear understanding of the provenance of each sub-category, demonstrating the alignment and integration of various frameworks in shaping the RAI Question Bank.
Detailed concept maps for each principle, including the relationships between concepts, are presented in the following sections.

\begin{comment}
\begin{figure*} [htb]
    \centering
    \includegraphics[width=\textwidth]{Fig/ConceptMap.pdf}
    \caption{The RAI concept map for RAI; It comprises 26 categories identified from seven AI frameworks.}
    \label{fig:concept map}
\end{figure*}
\end{comment}

\textit{Sub-question} refers to a group of questions associated with the sub-category. 
These questions are to identify potential risks during AI system development, ensuring comprehensive scrutiny and resolution of all potential risks.
For example, the question \textit{"Do you establish mechanisms that facilitate the system’s auditability?"} is to assess whether appropriate mechanisms are in place to enable auditing and review of the AI system's decision-making process. 
An unauditable system may hinder the identification of errors or biases, leading to negative consequences for stakeholders and society as a whole.

\textit{ID} in the figure represents a question ID which is used to uniquely identify each question within the question bank. 
There are two types of ID, such as internal ID and global ID.
Internal ID is employed for managing the questions internally, including tracking the source framework of each question. 
This is for maintaining mapping transparency and accuracy. 
All initially identified questions (a total of 382) have been assigned its internal ID.
The refined questions (245 in total) have been assigned its global ID, which is a universal reference number for external entities such as users and risk assessment systems. 

\textit{Level} is to assign the question level (1, 2, or 3) of the sub questions, indicating its importance and level of details within the risk category. 
The question levels are essential for facilitating tiered risk assessment, allowing for the effective use of the numerous questions available, as it may not be feasible to address all of them.
This accommodates the needs of different groups of stakeholders in the risk assessment process and provides a top-down approach that allows organizations to focus on the most important and high-level risks first.
The top-level question (i.e., level 1) is regarded as the fundamental question that provides a broad overview of the key risks.
These questions are primarily intended for high-level decision-makers such as C-levels or senior managers.

Subsequent questions are organized into second and third levels.
Level 2 questions are typically more specific than level 1 questions and provide additional detail and context.
These questions target managers and individuals (e.g., team members) with specific roles in overseeing AI-related projects.
Level 3 questions are even more detailed and enables users to further break down the key risks.
These questions are aimed at practitioners who require a detailed understanding of the technical and operational aspects of AI risk assessment. 
%\qinghua{how about mapping to stakeholders?}
The following shows the example questions at different levels.

\begin{itemize}
    \item \textbf{Level 1:} \textit{”Does the outcome result in something that all users can understand?”}, is an import and high-level question. However, it does not delve into specific considerations and risks in the system’s development.
    \item \textbf{Level 2:} An possible second-level question is \textit{"Do you design the AI system with interpretability in mind from the start?"}. This question assesses whether the project team considers "transparency by design" and manages potential design issues in the early development stages. As such, this question is suitable for management-level stakeholders rather than C-levels or senior managers.
    \item \textbf{Level 3:} \textit{”Do you research and try to use the simplest and most interpretable model possible for the AI system?”} can be used to further explore the topic if the development team has conducted sufficient research on interpretability during the model selection process.
\end{itemize}

\textit{Stage} refers to the specific point in the AI system lifecycle during where a particular question should be asked to ensure responsible and ethical development.
The AI system lifecycle consists of several critical stages, each with its unique focus and activities such as planning, requirements, design, implementation, testing, deployment and operation.
By organizing questions according to these stages, the RAI Question Bank ensures that ethical considerations are integrated throughout the entire AI system lifecycle, promoting a thorough and consistent approach to responsible AI development.

\textit{Source} represents the existing reference frameworks used by this study as depicted in Figure \ref{fig:QB}.
the RAI Question Bank initially included five AI frameworks such as \textit{NIST AI Risk Management Framework (NIST)}, \textit{EU Trustworthy AI Assessment list (EU)}, \textit{Canada Algorithmic Impact Assessment (AIA)}, \textit{NSW AI Assurance Framework (NSW)}, and \textit{Microsoft RAI Impact Assessment (MS)}, from different regions such as the US, Europe, Canada, Australia and a global company, respectively. 
The latest version has added \textit{the EU AI Act (EU Act)} and \textit{ISO/IEC AI management standard (42001:2023) (ISO)} to support compliance checking and provide expanded services to users. 

Table \ref{tab:QB summary} presents a summary of the RAI Question Bank. The table shows how the RAI Question Bank is organized into categories, sub-categories, questions, and the source frameworks.

\begin{table}[htb]
  \caption{The summary of the RAI Question Bank.}
  \label{tab:QB summary}
  \footnotesize
  \begin{tabular}{p{0.1\textwidth}p{0.15\textwidth}p{0.15\textwidth}p{0.15\textwidth}p{0.15\textwidth}}
    \hline
    Principle & \#Category & \#Sub-category & \#Sub-question & \#Source \\
\hline
\multicolumn{1}{l}{P1. HSE wellbeing} & \multicolumn{1}{r}{3} & \multicolumn{1}{r}{7} & \multicolumn{1}{r}{14} & \multicolumn{1}{r}{4} \\
\multicolumn{1}{l}{P2. Human-centred values} & \multicolumn{1}{r}{3} & \multicolumn{1}{r}{5} & \multicolumn{1}{r}{17} & \multicolumn{1}{r}{3} \\
\multicolumn{1}{l}{P3. Fairness} & \multicolumn{1}{r}{2} & \multicolumn{1}{r}{6} & \multicolumn{1}{r}{32} & \multicolumn{1}{r}{4} \\
\multicolumn{1}{l}{P4. Privacy/security} & \multicolumn{1}{r}{3} & \multicolumn{1}{r}{9} & \multicolumn{1}{r}{47} & \multicolumn{1}{r}{6} \\
\multicolumn{1}{l}{P5. Reliability/safety} & \multicolumn{1}{r}{5} & \multicolumn{1}{r}{11} & \multicolumn{1}{r}{42} & \multicolumn{1}{r}{6} \\
\multicolumn{1}{l}{P6. Transparency/explainability} & \multicolumn{1}{r}{3} & \multicolumn{1}{r}{9} & \multicolumn{1}{r}{32} & \multicolumn{1}{r}{4} \\
\multicolumn{1}{l}{P7. Contestability} & \multicolumn{1}{r}{2} & \multicolumn{1}{r}{4} & \multicolumn{1}{r}{4} & \multicolumn{1}{r}{2} \\
\multicolumn{1}{l}{P8. Accountability} & \multicolumn{1}{r}{5} & \multicolumn{1}{r}{14} & \multicolumn{1}{r}{57} & \multicolumn{1}{r}{6} \\
\hline
& \multicolumn{1}{r}{26} & \multicolumn{1}{r}{65} & \multicolumn{1}{r}{245} & \multicolumn{1}{r}{2-6} \\
\hline
\end{tabular}
\end{table}

\begin{comment} blocked due to page limit
\begin{figure*}[htb]
    \centering
    \includegraphics[width=\textwidth]{Fig/FrameworkPrinciple.pdf}
    \caption{The distribution of risk questions and the source frameworks across the eight AI ethics principles.}
    \label{fig:framework-principle}
\end{figure*}
\end{comment}

As shown in Table \ref{tab:QB summary}, the RAI Question Bank comprises 26 categories, 65 sub-categories, and 245 questions, derived from seven source frameworks. Each AI principle includes 2 to 5 categories with 4 to 57 questions to support AI risk assessment. \textit{Contestability} has the fewest questions, with only four, while \textit{Accountability} has the most, with 57 questions.

\begin{comment} blocked due to page limit
The source frameworks, including the EU AI Act and ISO standards, contribute to the distribution of the RAI Question Bank risk questions across the eight AI ethics principles. 
As depicted in Figure \ref{fig:framework-principle}, while most frameworks cover five or six principles, the ISO standard particularly concentrates on \textit{Accountability} as it primarily addresses management aspects.
\end{comment}

The following sections describe in detail how the RAI Question Bank is organized by the eight AI ethics principles. 

\subsection{Human, Societal, Environmental Wellbeing}
The rapid diffusion of AI has touched nearly every aspect of our lives. Many AI applications are developed using human-contributed data, directly interact with users (e.g., chatbots), or perform tasks on behalf of humans. However, the rise of AI is accompanied by concerns such as job displacement and harmful outcomes, both for individuals and society at large \cite{zhao2024rise, smuha2021beyond, zanzotto2019human}. Specifically, the widespread adoption of generative AI can introduce risks of toxicity, such as promoting hate, violence, or causing offense \cite{weidinger2021ethical}.
While the societal harms of AI are sometimes overlooked, its environmental impacts such as carbon emissions, energy consumption, and sustainable ecosystems have been widely discussed across various industries \cite{yigitcanlar2020contributions}.
These concerns underscore the critical challenge of designing AI systems that support human health and well-being \cite{van2023framework}.

\begin{figure*}[htb]
    \centering
    \includegraphics[width=\textwidth]{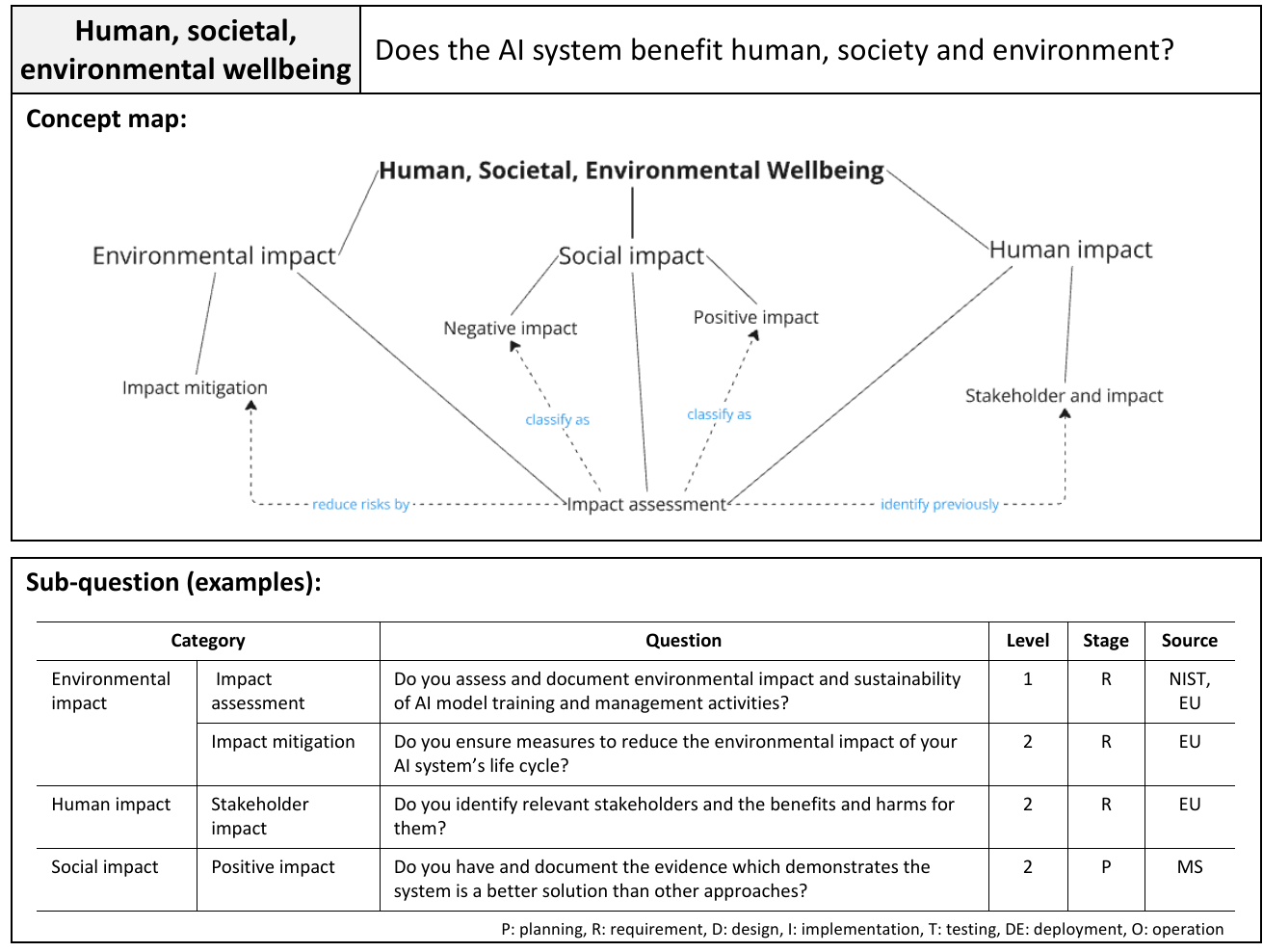}
    \caption{Human, Societal, Environmental Wellbeing.}
    \label{fig:P1}
\end{figure*}

The RAI Question Bank can be used to assess both positive and negative impacts of the AI system on human, society and the environment. 
The key question is defined as \textit{"Does the AI system benefit human, society and environment?"} (Figure \ref{fig:P1}).
There are a total of 14 sub-questions under the principle question to dive deeper and conduct practical-level risk assessments.

It consists of three key categories: environmental impact, social impact and human impact.
There are 2 - 3 sub categories under each category.
\textit{Impact assessment} has been identified as a key and common subcategory across all three risk categories.

As a top-level question for environmental impact assessment, we have identified a question, \textit{"Do you assess and document environmental impact and sustainability of AI model training and management activities?"} which is derived from \textit{NIST AI Risk Management Framework (NIST)} and \textit{EU Trustworthy AI Assessment list (EU)}.
Other questions under the same category can be used as subsequent questions to explore the relevant risks in further detail.
For example, \textit{Do you ensure measures to reduce the environmental impact of your AI system’s life cycle?} can follow the top-level question to assess how identified risks are managed.
This approach allows users to effectively navigate the questions and provides an appropriate order for addressing them.

Although the questions pertain to the same category, users may not need to use simultaneously. 
We have defined appropriate stages for each question, considering the AI development life cycle as mentioned in the previous section. 
Figure \ref{fig:P1} shows example questions and the proposed stages at which users should ask them.

\subsection{Human-centred Values}
With the increasing reliance on AI-based decision-making systems in recent years, there is a growing demand for ethical AI, which is often perceived as requiring human oversight of automation \cite{koulu2020human}. Although AI can significantly enhance performance compared to manual decision-making processes, over-reliance on these systems can result in negative consequences, particularly in complex or high-stakes situations where human oversight is essential \cite{buccinca2021trust}. Therefore, emphasizing human agency and oversight has become a key requirement for the effective implementation of AI applications \cite{koulu2020proceduralizing, enqvist2023human}.

\begin{figure*}[htb]
    \centering
    \includegraphics[width=\textwidth]{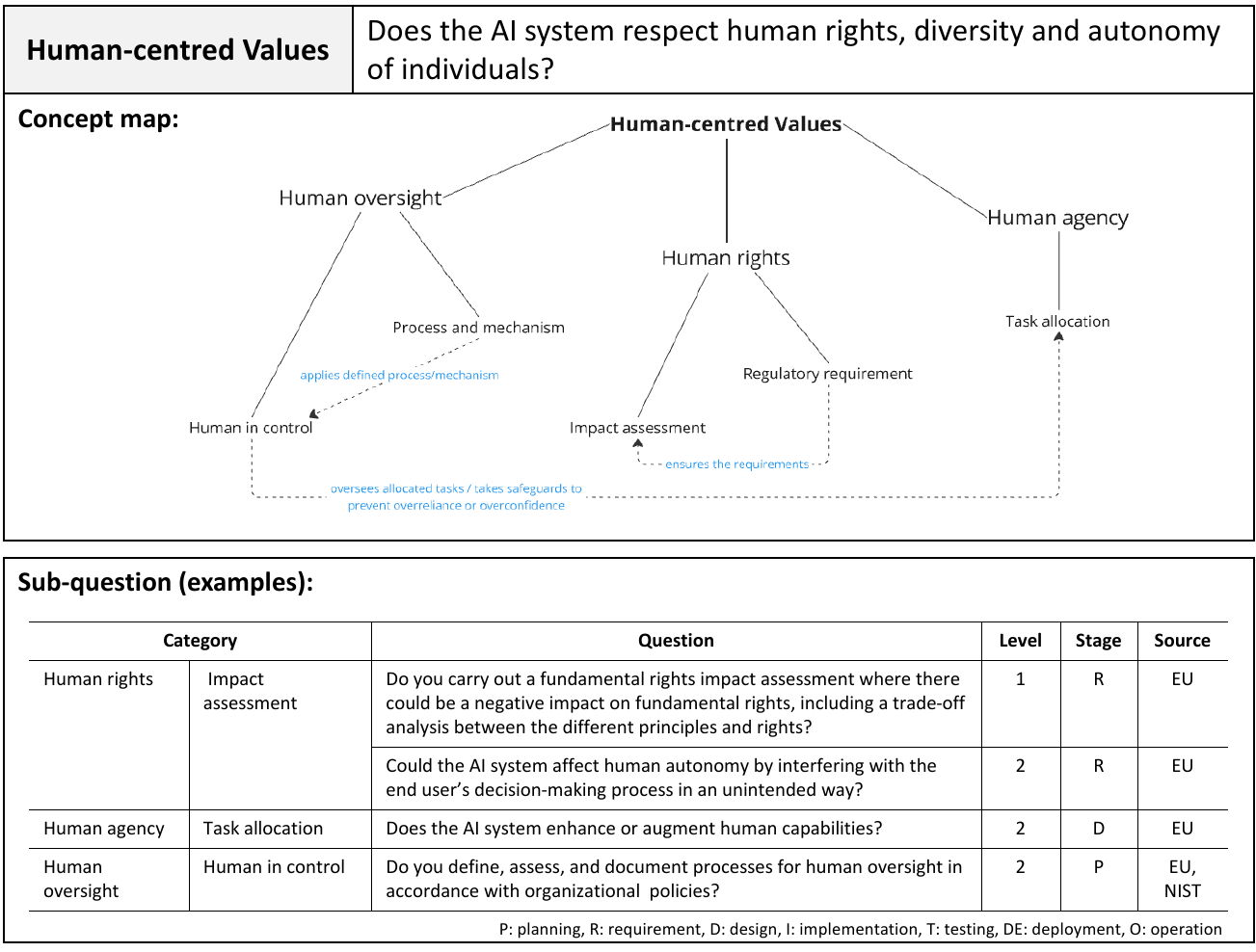}
    \caption{Human-centred Values.}
    \label{fig:P2}
\end{figure*}

To cope with these concerns, RAI should consider an appropriate level of human oversight and control to prevent AI risks, particularly those associated with human rights.
Under the key question, \textit{"Does the AI system respect human rights, diversity and autonomy of individuals?"}, the RAI Question Bank includes 17 sub-questions primarily based on three frameworks: NIST, EU and Microsoft.
The questions are categorized into three: \textit{human rights}, \textit{human oversight} and \textit{human agency} (Figure \ref{fig:P2}).

\textit{Human rights} is designed with five dedicated questions to identify risks related to regulatory requirements and ensure that the AI system respects human rights.
This category emphasizes the impact of technology design such as AI system on human autonomy \cite{calvo2020supporting}.
Human autonomy refers to the ability to willingly choose to be either dependent or independent \cite{deci2012self}.
Ensuring that AI systems uphold human autonomy is crucial for fostering trust and preventing unintended manipulation or coercion.

To maintain effective human oversight and intervention, there should be a clear delineation of responsibilities between human operators and AI systems \cite{cavalcante2023meaningful}.
\textit{Human agency} category comprises three questions and includes one sub-category, \textit{Task allocation}.
This assesses how the AI systems and human share tasks, whether the AI system enhances or augments human capabilities, and if there is an appropriate level of control over tasks allocated to AI systems.

\textit{Human oversight} includes nine sub-questions aimed at assessing AI risks stemming from insufficient human involvement, with processes and mechanisms in place to govern and control AI. 
Additionally, it recommends direct and immediate solutions to mitigate negative impacts, such as a stop button or procedure to safely abort operations when necessary, as commonly required by AI regulations such as the EU AI Act \cite{castets2022ex}.
We suggest that these questions should be used at the early stage (i.e., planing stage) of the AI lifecycle to establish a solid governance framework early on and support its implementation across the categories.

\subsection{Fairness}

\textit{Fairness} has attracted significant attention from researchers seeking to address the negative consequences of AI, such as unfair biases that can lead to discrimination or poor system performance \cite{benjamins2019responsible}.
AI systems should be inclusive, accessible, and free from unfair discrimination against individuals, communities, or groups \cite{alam2023developing}. Adopting such an approach promotes responsible AI design and deployment, minimizing bias and fostering inclusion and fairness.
Bias can emerge at various stages of AI systems, including in the data, models, or outputs. As AI systems learn patterns from input data to generate results, it is essential to address fairness throughout the entire AI life cycle \cite{chen2023ai}. This requires ongoing efforts to identify, monitor, and control bias at every stage.

\begin{figure*}[htb]
    \centering
    \includegraphics[width=\textwidth]{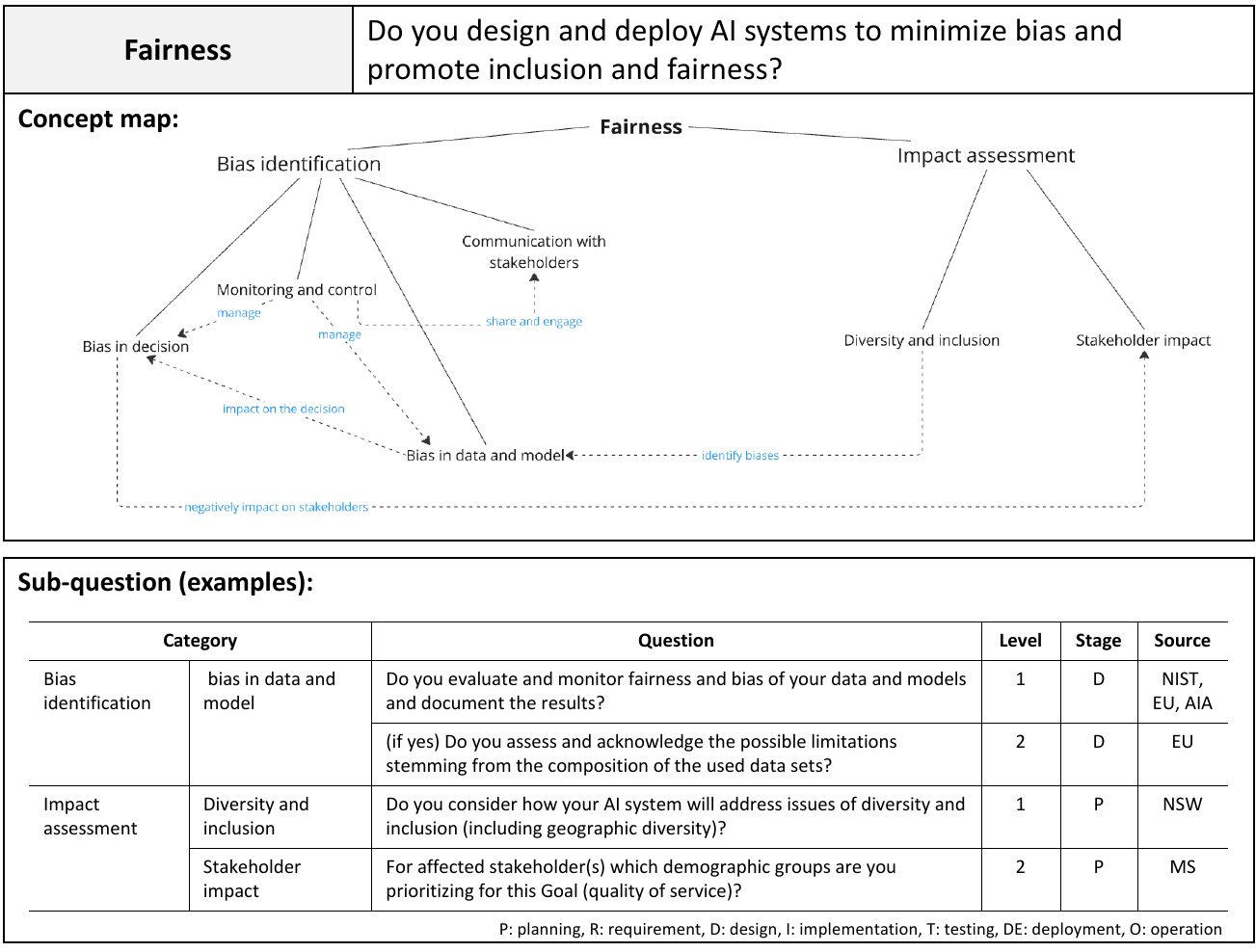}
    \caption{Fairness.}
    \label{fig:P3}
\end{figure*}

The RAI Question Bank includes 32 questions to thoroughly assess risks associated with this concern (Figure \ref{fig:P3}).

\textit{Bias identification} encompasses 12 questions to support identification of bias in training data and AI models such as gender bias, racial bias, age bias, geographic bias which can cause underrepresentation or overrepresentation.
There are also specific questions to identify any possible decision variability that can occur under the same conditions and to understand the potential impact of such variability on fundamental rights. 
These identified biases need to be comprehensively managed and transparently shared with stakeholders. 
However, this should follow a clear identification of potential stakeholders, including users and others indirectly affected.

\textit{Impact assessment} consists of 20 questions divided into two sub-categories: Diversity and Inclusion, and Stakeholder Impact.
\textit{Diversity and Inclusion} sub-category includes specific questions designed to ensure that the AI system properly addresses risks arising from the absence or lack of consideration for diverse group involvement, such as communities, minorities, and diverse teams, along with their feedback and data. Neglecting these considerations can lead to biases in the training data and the AI models, ultimately affecting the system's outputs.
On the other hand, \textit{Stakeholder Impact} sub-category aims to gain insights into the individuals or groups affected by the AI system's outputs. We have identified 12 dedicated questions for this category to understand the comprehensive range of stakeholders for different types of AI systems.

There are various techniques and metrics to support AI fairness. Disparate impact, equal opportunity, and statistical parity are commonly used to assess whether a system was designed without discriminatory intent, ensuring equal opportunities and statistically similar outcomes for all stakeholder groups \cite{garg2020fairness, mehrotra2022revisiting, ezzeldin2023fairfed}.
However, achieving fairness often involves trade-offs, as optimizing for one fairness metric can sometimes lead to reduced performance in another area or conflict with other ethical principles \cite{hooker2021moving}.

\subsection{Privacy and Security}

AI systems rely heavily on human-contributed data, raising significant concerns about personal information, sensitive data, privacy violations, and copyright issues \cite{verma2023copyright}.
Several privacy-preserving solutions, such as federated learning \cite{kairouz2021advances} and differential privacy \cite{dwork2006differential}, have been proposed, though no perfect solution exists. This highlights the importance of implementing privacy and security measures during the design phase—or earlier—and maintaining them throughout the entire life cycle to minimize technical debt \cite{cavoukian2021privacy, larrucea2021towards}.

\begin{figure*}[htb]
    \centering
    \includegraphics[width=\textwidth]{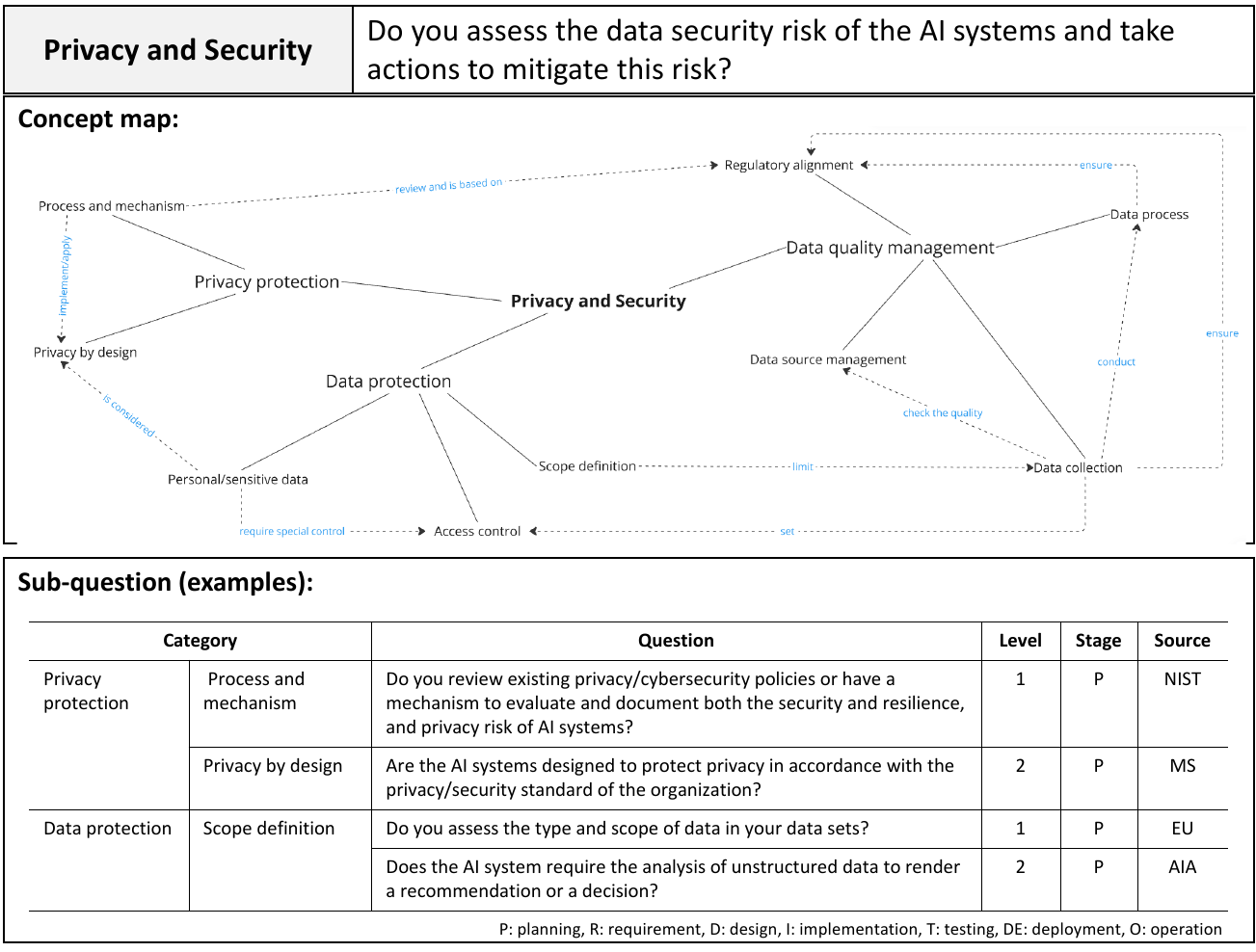}
    \caption{Privacy and security.}
    \label{fig:P4}
\end{figure*}

AI systems must ensure personal data privacy across their life cycle, including data traceability and protection against unauthorized access \cite{diaz2023connecting}. 
Figure \ref{fig:P4} illustrates the RAI concept map related to these concerns, along with key risk questions and example follow-up questions.
Privacy and security concerns can be assessed through 47 questions, categorized into three risk areas: privacy protection, data protection, and data quality management. 
These questions help ensure that AI systems are designed with privacy and security in mind from the early stages, safeguarding sensitive and copyrighted data throughout their operation.

\textit{Privacy protection} consists of 15 questions aimed at identifying risks associated with a company's privacy and security policies and their implementation in AI system design. This category emphasizes the need to review existing policies and regulations to ensure compliance with key requirements and recommendations.

\textit{Data protection} encompasses specific questions related to RAI practices regarding data definition, including input data types and scope for training AI models. The goal is to limit data collection and use for AI systems in compliance with relevant privacy policies and regulations. Special attention must be given to managing personal information and sensitive data when designing and operating AI systems, with rigorous access control and evaluation of existing data governance protocols.

\textit{Data quality management} helps users understand the risks in the data management process of AI, covering areas from data source management to data processing. This category includes questions to identify risks associated with the lack of regulatory alignment, ensuring comprehensive risk assessment in data handling for AI systems.

\subsection{Reliability and Safety}

\textit{Reliability and safety} are key components of a system's trustworthiness \cite{kuleshov2020addressing}. AI systems should be designed, developed, and tested with a focus on human safety, particularly in safety-critical domains such as healthcare, where system failures can directly cause patient harm \cite{anagnostou2022characteristics}.
Reliability and safety go beyond achieving high accuracy. According to the EU AI Act, high-risk AI systems must ensure comprehensive robustness, including appropriate accuracy metrics, resilience to errors, faults, or inconsistencies, risk management, and fail-safe mechanisms. These systems should also be tested against predefined metrics and probabilistic thresholds aligned with their intended purpose \cite{euaiact}. A holistic approach is essential to meeting these requirements.

\begin{figure*}[htb]
    \centering
    \includegraphics[width=\textwidth]{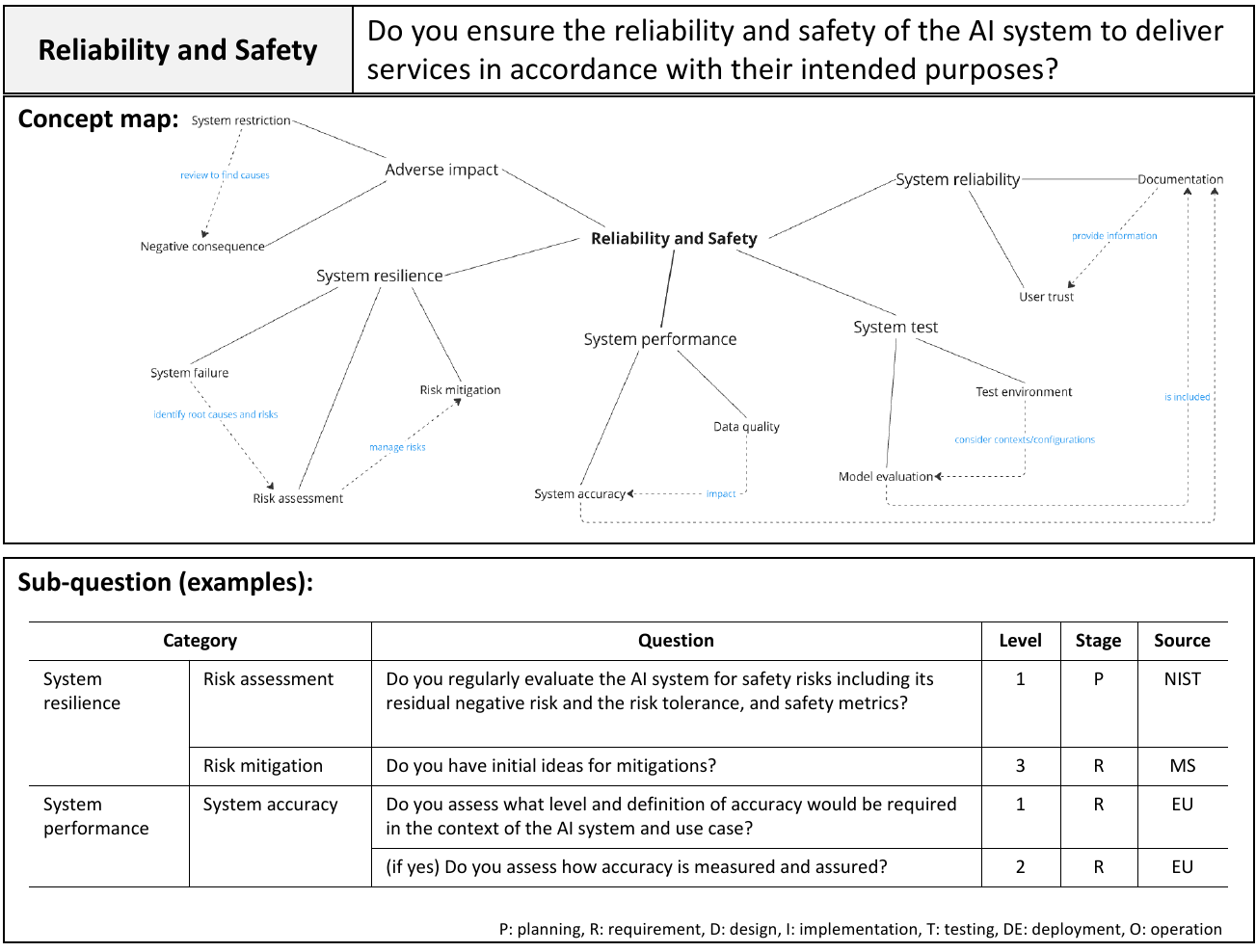}
    \caption{Reliability and safety.}
    \label{fig:P5}
\end{figure*}

We have identified five categories for these concerns: system performance, system test, system reliability, system resilience, and adverse impact (Figure \ref{fig:P5}). Together, these categories encompass a total of 42 risk questions designed to thoroughly assess and ensure the reliability and safety of AI systems.

\textit{System performance} includes questions that evaluate the overall performance of the AI system, ensuring it meets the expected benchmarks and operates efficiently under various conditions. This category focuses on metrics such as accuracy and data quality.

\textit{System test} covers questions related to the testing methodologies and environment employed to validate the AI system. It focuses on model evaluation considering the comprehensiveness of the test cases, the coverage of different scenarios (including edge cases), and the robustness of the testing processes. This category ensures that the system has been rigorously tested before deployment. The test results and any issues and limitations should be documented and shared with stakeholders.

\textit{System reliability} involves questions that assess the dependability of the AI system over time. This category looks into the system's ability to perform consistently without failure, the measures taken to maintain reliability with proper documentation to improve user trust.

\textit{System resilience} includes questions aimed at understanding the AI system's ability to withstand and recover from disruptions or adverse conditions. This category examines the system's fault tolerance, redundancy mechanisms, and recovery procedures to ensure continuous operation even in the face of unexpected challenges. System failures should be addressed managed by risk assessment and risk mitigation practices.

\textit{Adverse impact} encompasses questions that identify and mitigate potential negative impacts of the AI system on users and other stakeholders. This category focuses on evaluating the system's behavior in critical situations, its compliance with ethical standards, system restriction and the measures taken to prevent harm or unintended consequences.

Collectively, these 42 risk questions provide a comprehensive framework to ensure that AI systems are reliable, safe, and capable of delivering consistent performance while minimizing potential risks and adverse impacts.

\subsection{Transparency and Explainability}

There is broad consensus that AI systems, particularly deep learning models, are often opaque and difficult for humans to understand—commonly referred to as the black box problem \cite{von2021transparency}. This lack of transparency has contributed to diminished user trust, driving increased demand for transparent and explainable AI in the RAI space \cite{arrieta2020explainable}.
As a result, most AI governance frameworks and regulations now include \textit{Transparency and Explainability} as essential principles for developing and deploying trustworthy AI systems \cite{euaiact, eutrustworthyai, oecdaiprinciple}.

Traditionally, AI transparency focused on algorithms, aiming to demonstrate how decisions were made \cite{haresamudram2023three}. However, recent frameworks call for more comprehensive information about AI systems, including specifications such as the intended purpose, data sources, system performance, required training resources (e.g., computing power and time), and limitations (e.g., unmitigated risks) \cite{euaiact}.
Additionally, interpretability should be prioritized as a key design element to ensure impartiality, robustness, and meaningful insights from system variables \cite{arrieta2020explainable}.

\begin{figure*}[htb]
    \centering
    \includegraphics[width=\textwidth]{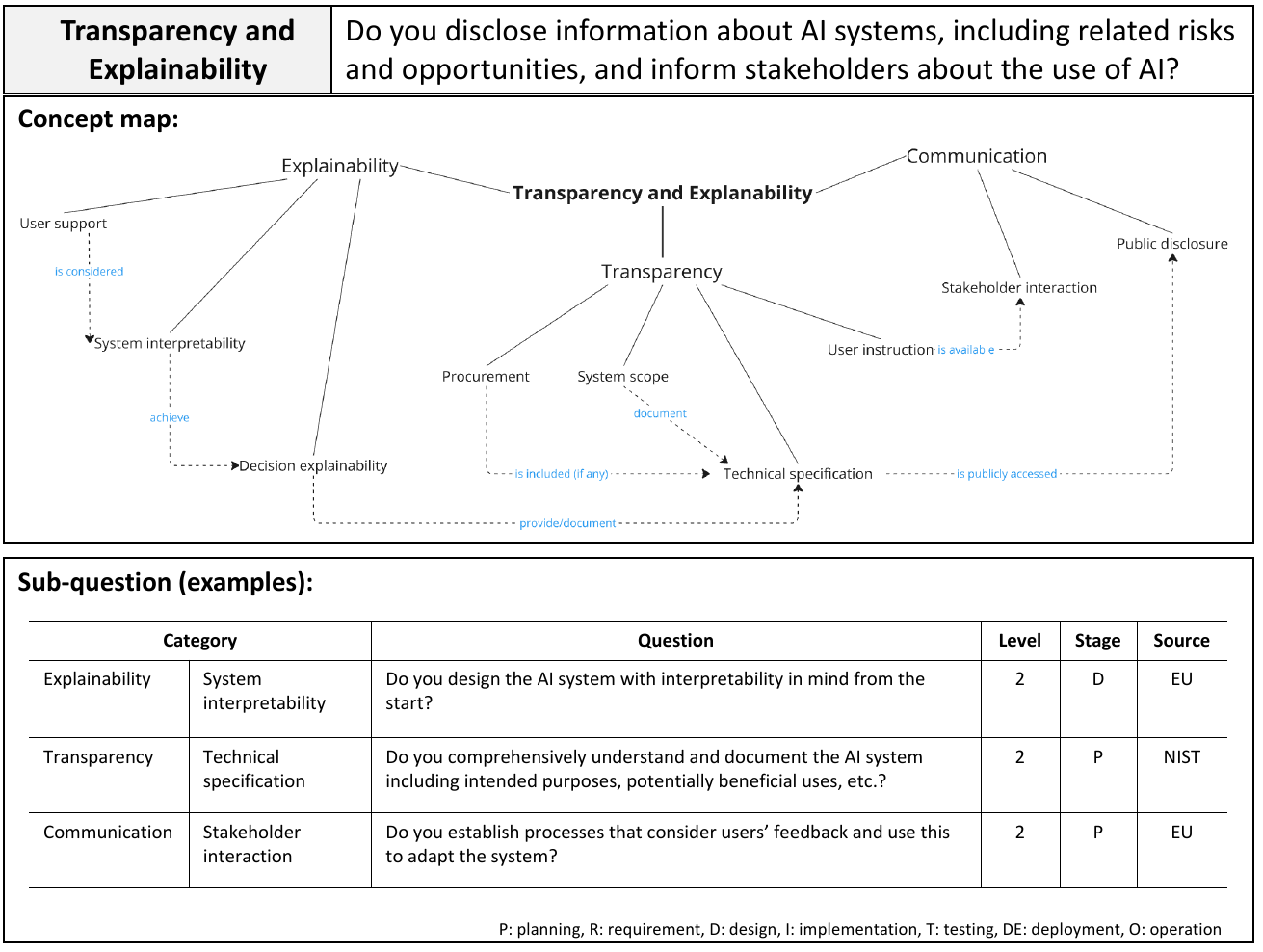}
    \caption{Transparency and explainability.}
    \label{fig:P6}
\end{figure*}

To address these concerns, the RAI Question Bank includes 32 questions for three risk categories (Figure \ref{fig:P6}): \textit{Transparency}, \textit{Explainability}, and \textit{Communication}. This principle specifically underscores the responsibility of the AI model and system providers towards the users to ensure that AI systems are transparent, understandable, and effectively communicated.

\textit{Transparency} involves questions that assess the clarity and openness of the AI system’s operations and decision-making processes. This category examines whether users and stakeholders have access to sufficient information about how the AI system functions, the data it uses, and the algorithms it employs through technical specifications. 
It emphasize that all relevant stakeholders should be able to understand the system scope, goals, limitations and potential risks of the AI systems.

\textit{Explainability} includes questions aimed at determining how well the AI system’s decisions and behaviors can be interpreted and understood by users. This category focuses on the ability of the AI system to provide clear, understandable explanations for its actions and outcomes, helping users trust and effectively use the system.

\textit{Communication} encompasses questions that evaluate the effectiveness of the information exchange between the AI system and stakeholders including users. 
Providing information to non-technical audiences may require managing ambiguities, but the information should be accessible for society, policy makers and any other stakeholders. 
 \cite{arrieta2020explainable}. 
This category ensures that the system provides timely, accurate, and useful information to them, facilitating informed decision-making and fostering a positive user experience.

By addressing these categories, AI systems are designed and deployed with a strong emphasis on user-centric principles, thereby promoting trust, accountability, and the effective use of AI technologies.

\subsection{Contestability}

Contestability refers to the ability to challenge and question AI decisions, ensuring users and affected parties can seek redress or have decisions reviewed \cite{lyons2021fair}. 
Most existing RAI frameworks do not explicitly address or include this principle. We have observed that the Canada AIA and Australia NSW frameworks include a few questions related to this principle. Other frameworks tend to address this implicitly under their \textit{Transparency} or \textit{Fairness} principles.
This issue has resulted in a lack of guidance on designing AI systems that support contestation.

In human decision-making, it is generally easy to obtain the reasoning behind a decision from the decision-maker through internal or external reviews and complaints mechanisms. However, AI-based decision-making is inherently complex and often operates at scale, making it difficult for stakeholders to access the decision process and raise objections \cite{landau2024challenging}.
It underscores that AI contestability should be incorporated early in the development process, with mechanisms embedded within the AI system through an interface.
This \textit{Contestability by Design} concept can also help AI systems achieve fairness by making them contestable for all stakeholders \cite{von2024designing}.

In line with these considerations, we identified four questions to support this principle, focusing on human interfaces and the right to appeal. 
These questions address mechanisms for users to contest AI decisions, the processes in place for reviewing contested decisions, and the transparency of these processes. 
By incorporating contestability measures, AI systems promote accountability and provide users with fair means to address potential biases or errors in AI outputs.

\begin{figure*}[htb]
    \centering
    \includegraphics[width=\textwidth]{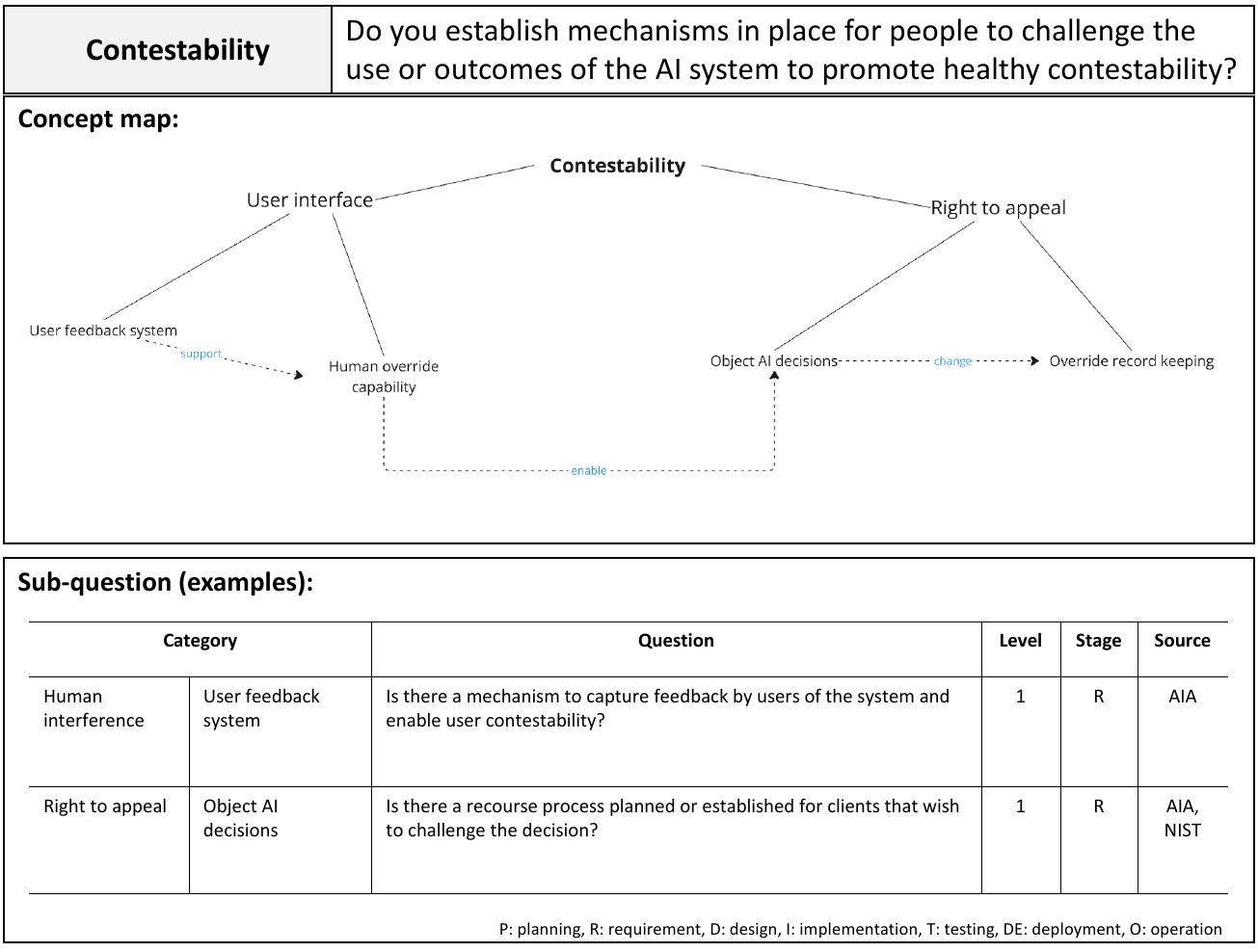}
    \caption{Contestability.}
    \label{fig:P7}
\end{figure*}

Figure \ref{fig:P7} illustrates the concept map and associated risk questions for user contestability, highlighting the mechanisms for challenging AI decisions and ensuring fair review processes.

\subsection{Accountability}

\textit{Accountability} in the RAI space is defined as the principle that AI actors should be held responsible for the proper functioning of AI systems \cite{oecdaiprinciple}.
It extends beyond merely defining roles and responsibilities of AI actors to include associated tasks and activities, such as ensuring traceability (across datasets, processes, and decisions) and implementing essential management frameworks, including quality, risk, leadership, and competency management \cite{london2024accountability, metwally2024thinking, clarke2019principles}.

Accountability challenges include unclear responsibility for AI outcomes and the absence of mechanisms to address unintended consequences \cite{khan2024responsible}.
Accountability is closely tied to other RAI principles, such as Fairness, Privacy, and Transparency, as it provides fundamental management frameworks that support their implementation.
As it plays a pivotal role, the lack or absence of an appropriate level of accountability in RAI can critically impact trust in AI systems and lead to misuse, unethical practices, or harmful outcomes.

\begin{figure*}[htb]
    \centering
    \includegraphics[width=\textwidth]{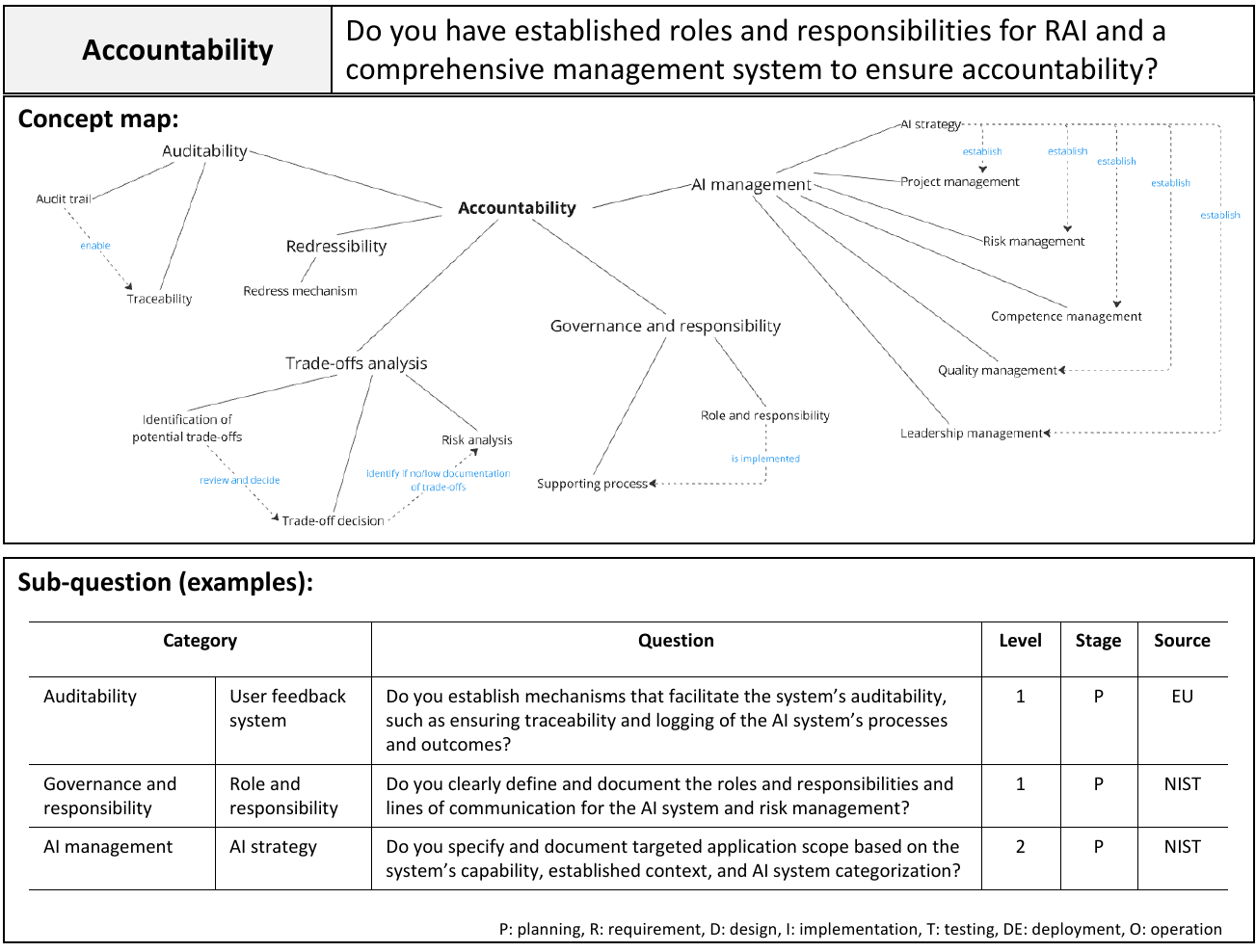}
    \caption{Accountability.}
    \label{fig:P8}
\end{figure*}

To address these risks, the RAI Question Bank includes 57 questions distributed into five categories: \textit{Auditability}, \textit{Trade-offs analysis}, \textit{Redressibility}, \textit{Accountability framework}, and \textit{AI management} (Figure \ref{fig:P8}).
This principle encompasses a more extensive set of questions compared to other principles due to recent additions aligned with emerging AI policies such as the EU AI Act and ISO AI management. 
The heightened focus on comprehensive AI quality and management within these regulations has generated such additional accountability-related practices and risk questions.

\textit{Auditability} ensures that AI systems are designed with mechanisms that allow for comprehensive audits, enabling stakeholders to trace and verify decisions made by the AI \cite{li2024making}. This category includes 11 questions that assess the ability to monitor and review AI operations and outcomes, ensuring transparency and trustworthiness.

\textit{Trade-offs analysis} focuses on the decision-making processes within AI systems, particularly how they balance competing objectives and values \cite{kumar2024balancing}. 
This includes questions that identify relevant interests and values implicated by the AI system, potential trade-offs between them, and the consideration of ethical implications and potential conflicts between different goals. 
This ensures that AI systems make balanced and fair decisions while also aiming to understand potential risks associated with the lack or absence of management of trade-offs.
An example of trade-off analysis is an accuracy-energy experiment, which demonstrates that in some cases, a small reduction in accuracy can result in significant energy savings \cite{brownlee2021exploring}.
Another important trade-off for AI systems is between data quality and data availability.

\textit{Redressibility} addresses the mechanisms in place for correcting or mitigating harm caused by AI systems \cite{gardner2022responsibility}. It includes two questions that assess the existence and effectiveness of processes for users to seek redress in case of any harm or adverse impact and provide information to the users (or third parties) about opportunities for redress.

\textit{Accountability framework} is pivotal to ensuring that there are clear roles and responsibilities defined for overseeing RAI practices. This category includes three questions that assess if there is clearly assigned accountability and supporting processes to implement these accountability practices.

\textit{AI management} covers the broader management practices surrounding AI development and deployment. It includes 40 questions that evaluate the ongoing monitoring, updating, and governance of AI systems, ensuring that they remain aligned with ethical principles and organizational goals throughout their lifecycle. 
Specifically, 22 questions are derived from the ISO AI management standard, addressing comprehensive AI management practices including AI project management, risk management, competency management, and leadership management. 

\section{Case Study} \label{sec:case study}

\subsection{RAI Risk Assessment in Scientific Research} \label{sec:case study_1}

\textbf{Case study overview.}
We have evaluated the ethical risks associated with eight AI projects, designated PR1 to PR8, within a national research organization in Australia. These projects primarily focus on scientific research and incorporate AI as a component of broader research initiatives. Each project has distinct goals and operates across various sectors such as food security and quality, health and well-being, and sustainable energy. Given that these projects are in their early stages, they provide a valuable opportunity to integrate best practices in responsible AI for future development.
The hierarchy of the project teams was relatively flat, consisting of a \textit{Project Lead} who manages the entire research project, multiple \textit{Work Package Leads} who oversee sections of the research project, and \textit{Researchers} who report to the Work Package Leads.

\textbf{Preparation— Development of a risk register template.}
Prior to conducting the risk assessment, we developed a dedicated risk register template that includes essential risk assessment components such as risk ID, risk category, risk description, risk causes, existing mitigation measures, risk owners, and interview questions. 
The RAI Question Bank was integral to this template, serving as the foundation for developing risk assessment categories, different assessment levels, and interview questions. The risk assessment register adhered to the risk categorization which includes 26 themes (see Figure ~\ref{fig:concept map}). Additionally, it utilized the three question levels and corresponding questions from the RAI Question Bank to formulate interview questions for each of these themes.

\textbf{Risk assessment— Engagement with project teams.}
The first round of risk assessment sessions was scheduled between January and February 2023. Each session included a 1.5-hour interview, with 10 minutes allocated for introduction and wrap-up. The participants included a risk assessor, 1-2 observers, and 1-4 interviewees. Project leads of all eight projects participated in the interviews, and in most cases, the work package leads also participated. In a few instances, researchers participated as interviewees as well.

In the first round, we focused on the high-level (level 1 and 2) questions of the RAI Question Bank for each AI principle. The interviews were recorded and transcribed using a paid service, and the transcriptions were rechecked for accuracy. Following the initial data analysis, a second round of interviews was conducted a few months later. It aimed to explore the identified risks in greater depth, focusing primarily on the level 3 questions from the RAI Question Bank.
%The collected interview data were analyzed to identify and assess the risks associated with AI ethics in these eight projects. For risk level calculations, we used a well-known 3x3 risk assessment matrix and an impact and probability equation~\cite{kovacevic2019application}.   
%\harsha{should we add the detailed description of how the final calculations were done from different sub categories to main category?}
%\sunny{I don't think so. Will see reviewers' comments later.}
%The risks were scored using the enterprise risk framework.

\begin{comment}
\textbf{Risk assessment results summary.} 
%\harsha{not sure whether this is needed}
Accountability emerges as the principle with the highest combined medium to high risk. This is mainly to do with missing a proper responsible distribution documentation and lack of traceability within projects. \harsha{will add a summary of the other results.}

%We have found that specific projects (e.g., PR4 and PR8) exhibit high risks related to privacy, reliability, and transparency, while the majority of projects have low risks in terms of contestability.    
\end{comment}

\textbf{Results— Effectiveness and feasibility.}
The question bank's structured approach, combining high-level principles with the different-level questions, has proven to be effective in uncovering potential risks across diverse AI projects. %By mapping questions to different development phases and incorporating feedback from users, we have ensured that the question bank addresses a broad range of risk factors and provides actionable insights.
The structured questions prompted interviewees to consider different aspects of their projects that they had not previously contemplated.
%We analyzed the projects to understand their contexts.
For instance, one of the Project Leads in PR3, which focuses on AI for drug discovery, had not considered the risks associated with third-party university agreements on intellectual property ownership, particularly regarding sharing chemical formulas stored in university repositories, until prompted by the RAI Question Bank questions. 
Through the first round of interviews, we identified \textit{Accountability} emerges as the principle with the highest combined medium to high risk. This is mainly to do with missing a proper responsible distribution documentation and lack of traceability within projects. 
\textit{Transparency and explainability} also pose significant risks, primarily due to insufficient clarity in explaining AI outcomes. %While accuracy is crucial in scientific discovery, it can result in less understandable results. Implementing additional measures is advised to ensure that stakeholders have a clear understanding of AI outcomes, which is vital for informed decision-making.
Conversely, \textit{Human-centred values}, \textit{Human, societal and environmental wellbeing} and \textit{Contestability} are recognized as relatively low-risk principles. These areas have demonstrated better alignment with ethical practices and project requirements, suggesting fewer immediate concerns.
The key findings and risks for each project were further discussed in the second round of interviews, confirming that the RAI Question Bank effectively highlights critical risk areas and provides a solid foundation for risk management. 

The feasibility of using the RAI Question Bank was demonstrated through its structured approach. It provided a comprehensive framework that facilitated the exploration of diverse risk areas, ensuring both breadth (covering the eight AI principles) and depth (addressing multiple question levels).
This structure enabled the efficient creation of a risk register template and formulation of interview questions. We analyzed project contexts and selected 3 to 5 questions per AI principle to guide the interviews.
The risk assessments incorporated the risk register template, a standard 3x3 risk matrix, and calculations for impact and probability to determine risk levels. By applying the question bank across eight projects, we confirmed its value as a focal resource, showing that it integrates smoothly into traditional risk workflows without requiring significant adjustments.

\textbf{Discussion— Practical challenges in real-world contexts.}
Durging this case study, several practical challenges were encountered when using the RAI Question Bank to assess ethical AI risk in real-world projects. Firstly, the stage of the project significantly influences the risk assessment process. For example, projects at the inception phase may have a limited understanding of model accuracy, which impacts the evaluation of the principle of reliability and safety under the sub category - system performance (see Figure~\ref{fig:P5}).
Secondly, the effectiveness of the interviews depends heavily on the interviewees. In our context, the flat hierarchy meant that project leads and work package leads, who were also researchers, had comprehensive knowledge of their projects. Therefore, they were able to answer all the question from level 1 to 3. However, we acknowledge that this may not be the case in many commercial or open-source collaborative projects, where individuals may only have knowledge specific to their roles.

\textbf{Feedback and improvement.} Following the risk assessment process, stakeholders provided valuable feedback. One key point was that the number of questions in the RAI Question Bank sometimes seemed excessive relative to the level of completeness needed for the assessment. While the entire question bank may useful for ethical AI assessments in mission-critical or life-critical systems, a subset of questions could be sufficient for systems with lower risk levels.

Another point of feedback was the relevance of the questions. Although the risk questions were generally useful for identifying risks, not every question was applicable at every stage of the project. This was recognized as a manageable challenge but highlighted the need for improvement.

%Based on this feedback, stakeholders recommended enhancing the RAI Question Bank structure by incorporating a dimension that accounts for the development stage of the project. Tailoring questions to the specific stage of each project would enable more accurate and contextually relevant risk identification. This enhancement is expected to improve the applicability, efficiency, and effectiveness of the RAI Question Bank, contributing to more robust and comprehensive AI risk management practices.

Based on this feedback, we reviewed and refined the question bank, mapping questions to different development phases. This enhancement enables a more efficient and effective risk identification process, aligning questions with each project's specific stage. By tailoring questions to the development stage, potential risks can be identified more accurately and in a manner more applicable to the project's context. This not only enhances the relevance and usefulness of the question bank but also optimizes resource allocation and efforts in assessing AI risks.
This refinement addresses the challenge of assessing risk levels in early-stage projects with limited model accuracy insights and minimal real-world user engagement. By aligning questions with specific development stages, we ensure that the questions are more applicable and actionable throughout the project life-cycle. Additionally, this approach mitigates the impact of time constraints by prioritizing the most relevant ethical principles based on the project's current phase, ultimately leading to a more efficient and focused risk assessment process.

\subsection{RAI Deep Dive Assessment for Investors} \label{sec:case study_2}

\textbf{Case study overview.}
We utilized the RAI Question Bank for the integration of Environment, Social and Governance (ESG) and AI for investors.
This project aimed to develop the ESG-AI investor framework \cite{ESGAI}, which was implemented from April 2023 to April 2024. 
Collaborative research \cite{ward1982collaborative} was applied for this project, allowing researchers and practitioners to work together in developing the framework. 
The practitioners involved in this project included not only investors as potential users, but also senior design experts and industrial experts in both AI and ESG.
% The following is blocked due to page limit.
\begin{comment}
The following outlines the key roles and responsibilities defined in the project.

\begin{itemize}
    \item AI researcher: Provide insights into AI principles, technologies, and responsible AI practices; Drive the framework and toolkit development.
    \item ESG expert/investor: Share perspectives on investment strategies, ESG considerations and assessment needs.
    \item Design expert: Ensure the framework's usability from a user-centered 
    design perspective.
    \item Industrial expert: Contribute industry-specific knowledge and practical insights while balancing out different perspective from AI researchers and investors.
    \item Project manager: Coordinate the project activities and kept the research team on schedule.
\end{itemize}

The main creators of the RAI Question Bank were involved in this project as AI researchers and played a crucial role in developing the ESG-AI framework by applying the RAI Question Bank.
\end{comment}

\textbf{The ESG-AI framework and the application of the RAI Question Bank.}
The ESG-AI investor framework comprises three components that collectively form the backbone of the framework. These components include an in-depth review of AI use cases for industry sectors, an evaluation of RAI governance indicators, and RAI deep dive assessment for a meticulous examination of RAI principles. 
These elements provide a holistic approach to assessing and mitigating the risks associated with AI deployment, fostering responsible AI practices across diverse sectors.

The RAI deep dive assessment is \textit{primarily based on the RAI Question Bank}.
%, which incorporates insights and standards from key regulatory bodies, standard organizations, and stakeholder groups, including the EU AI Act, NIST AI Risk Management Framework, ISO AI Standard (ISO/IEC 42001), and other industry AI risk frameworks.
It has a similar structure to the RAI Question Bank and is a specific subset tailored for ESG integration. This version includes the eight AI ethics principles as overarching categories, each with eight lead questions and corresponding sub-questions. Initially, we selected questions from the RAI Question Bank that aligned with the 12 ESG topics identified by participating ESG experts: three environmental topics (carbon emissions, resource efficiency, and ecosystem impact), six social topics (diversity, equity, and inclusion; human rights and labor management; customer and community relations; data privacy and cybersecurity; and health and safety), and three governance factors (board and management; policy; and disclosure and reporting).

Through collaboration with ESG experts, we reviewed and tailored these questions through multiple iterations to suit the project's specific needs. The final version of the RAI deep dive assessment includes 42 sub-questions across the eight AI ethics principles. Additionally, we incorporated RAI metrics to support investors' assessment and encourage company transparency and public disclosure. 
%address Liming's comment
As discussed earlier, these metrics are specifically included to inject quality aspects into the questions and provide evidence beyond minimal yes/no answers.
Table \ref{tab:deep dive_accountability} shows the example questions and metrics selected for Accountability.

\begin{table}[htb]
    \centering
    \footnotesize
    \begin{tabular}{p{0.15\textwidth}p{0.18\textwidth}p{0.35\textwidth}p{0.22\textwidth}}
    \hline
    Key question & Indicator & Sub-question & Metric \\
    \hline
    \multirow{4}{0.15\textwidth}{Does the company have designated responsibility for AI and RAI within the organisation?} & Risk management & Does the company establish methods and metrics to quantify and measure the risks associated with its AI systems? & Number of AI risk metrics (e.g., risk exposure index, risk severity score) \\
     & AI incident management & Does the company have a clear reporting system or process in place for serious AI incidents to inform external stakeholders (e.g., market surveillance authorities, communities) beyond the company? & Number of AI incidents informed to external stakeholders \\
      & Accountability framework & Does the company have an accountability framework to ensure that AI related roles and responsibilities are clearly defined? & Percentage of defined AI roles and responsibilities \\
    \hline
    \end{tabular}
    \caption{Example questions and metrics (Accountability).}
    \label{tab:deep dive_accountability}
    % \normalfont
\end{table}

\textbf{User feedback and impact on the community.}
During the project, we conducted multiple workshops with investors (i.e., potential users) and senior industrial experts, engaging in \textit{iterative testing and improvements}. These dedicated participants contributed to the ongoing review process, providing valuable feedback on the framework's usability, particularly the risk assessment questions from the user perspective. 
Accordingly, user feedback primarily addressed the readability and usability of the questions and metrics in real-world scenarios, as well as the clarity of the guidance for each assessment item.
Feedback included recommendations such as:

\textit{Restructuring the question bank and adding a question for each AI ethics principle:} This change aimed to provide high-level insights and a better understanding of RAI practices and the risks associated with each principle. The previous version of the RAI Question Bank did not include principle-specific questions; however, based on this feedback, we modified the structure. The current version now includes principle questions and corresponding sub-questions.
    
\textit{Improving the clarity of questions and metrics that might be difficult for general users to understand:} In response, we identified and revised questions containing technical or specialized terms to enhance their quality and ensure they are more accessible. We also revised the metrics, including their descriptions, to provide clearer guidance and improve the overall assessment process. These changes aim to ensure that the questions and metrics are not only comprehensible to a wider audience but also maintain a high standard of quality and relevance.

\textit{Instructions for users on how to evaluate the answers of examinees:} This aims to enhance risk assessors' capability in accurately judging responses, ensuring consistent and effective assessments. These guidelines can help assessors better understand the criteria for evaluation, leading to more reliable and insightful risk assessments. 
%address Liming's comment
They assist users in addressing the quality aspects of questions by incorporating relevant metrics, beyond relying on simple yes/no answers. We have added specialized instructions for users into the framework, providing guidance on evaluating the quality of evidence and the context of responses. 
Due to the large scale of the RAI Question Bank, this approach has not yet been applied to the entire question set. We recognize the importance of these enhancements and will consider and discuss further improvements for future iterations.

This iterative process ensured that the risk questions in the framework was refined to meet the needs of its users effectively.
As a result, the RAI Question Bank-based questions was recognized as a highly useful and practical tool for AI and ESG risk assessment for investors.
The framework was publicly released in April 2023. 
Within the first 24 hours, it received coverage from 23 media outlets.  
In the first week, over 1,000 people downloaded our final report from the website, and around 100 downloaded the framework toolkit, despite the gated website requiring personal information. 
The framework also gained significant attention on social media, where it was praised for providing valuable insights into addressing key challenges faced by practitioners and for going beyond a principles-only approach to implementing RAI.

\section{Discussion} \label{sec:discussion}

\begin{comment}
\begin{itemize}
    \item Case study findings, feedbacks from projects (AI4M projects/Alphinity project)
    \item Any practical implications, challenges, limitations 
          Like clay.....tailor/customize....context/needs.... => projects...., Mention use case: QB for LLMs, foundation model.....compliance check....
    \item Scoring and navigating structure (?) for implementation
    \item Threats to validity
\end{itemize}    
\end{comment}

In this section, %we discuss important findings and feedback identified through the case studies, which serve as key inputs for the improvement and extension of the RAI Question Bank in future iterations. 
we demonstrate practical uses of the RAI Question Bank, which may be of interest to practitioners aiming to develop and use AI ethically.
Specifically, we demonstrate how the RAI Question Bank can help companies examine their RAI practices to comply with key AI regulations and standards such as the EU AI Act.
We also discuss the application of the RAI Question Bank for \textit{foundation model-based AI agents}, given their widespread adoption in this area \cite{masterman2024landscape}.

%\subsection{Key findings and feedback from the case studies}

%\sunny{move to the previous section?}

\subsection{Compliance with Regulations} \label{sec:complaince}

In recent years, many countries have been working to develop AI bills and regulations to ensure the secure use of AI.
For example, \textit{the EU AI Act} has become the first comprehensive regulation on AI in March 2024.
In the US, \textit{AI bill of rights} \cite{uswhitehouseblueprint} developed by the White House Office of Science and Technology Policy (OSTP) in 2022, serves as a blueprint regarded as a voluntary regulation.
\textit{Bill C-27} \cite{canadabillc27} in Canada, also known as the Digital Charter Implementation Act, is a proposed law currently making its way through the Canadian Parliament. 
In South Korea, \textit{Bill on AI Liability} was suggested in March 2023. It has not yet been passed into law. 
The Australian government has released the \textit{Voluntary AI Safety Standard} \cite{ausaisafety} and \textit{National framework for the assurance of AI in government} \cite{ausnationalframework} in 2024.
These regulations commonly include specific requirements for high-risk AI as it may pose significant risks to the health and safety or fundamental rights of natural persons (e.g., credit scoring \cite{langenbucher2022responsible} and facial recognition \cite{almeida2022ethics}). Companies developing and using AI are concerning about compliance with current and upcoming regulations \cite{perera2024achieving}.
To support this, we illustrate how practitioners can utilize the RAI Question Bank to examine their RAI practices against the legal requirements identified in the EU AI Act, based on our mapping approach depicted below.

\textbf{1) Identify legal requirements from the EU AI Act.}
We conducted a thorough review of the EU AI Act to identify the specific legal requirements that apply to high-risk AI systems. 
This involved analyzing the text of the Act to extract all relevant legal requirements that developers of high-risk AI systems must comply with. 
Through this process, we identified 21 key requirements that address various aspects of AI ethics, including transparency, accountability, fairness, and privacy.
These requirements are designed to ensure that high-risk AI systems do not pose significant risks to the health, safety, or fundamental rights of individuals. 
They include provisions for proper documentation, risk management, human oversight, data governance, and more. 
This step is crucial for understanding the regulatory landscape and establishing a foundation for systematic risk assessment and management of high-risk AI systems.
Table \ref{tab:high-risk requirement} presents 10 example requirements that we identified for high-risk AI systems.

\begin{table}[htb]
    \centering
    \footnotesize
    \begin{tabular}{p{0.23\textwidth}p{0.05\textwidth}p{0.65\textwidth}}
    \hline
    Category & ID & Requirement \\ 
    \hline
    Risk management & E01 & Risk management system shall be established, implemented, documented and maintained. \\
    Data governance and management & E02 & For datasets, should consider relevant data preparation, prior quality assessment, examination (e.g., biases), possible data gaps or shortcoming. \\
    Technical specification & E03 & General description (e.g., intended purpose) and  detailed description (e.g., design specification, architecture). \\
    Record keeping & E04 & Record keeping of all relevant documentation and information for traceability. \\
    Transparency & E05 & User instructions including provider details, capabilities and limitations, purposes, performance, specifications, etc. \\
    Transparency & E06 & If the AI system is interacting with humans, the users should be informed that they are interacting with an AI system and the content has been artificially generated or manipulated. \\
    Human oversight & E07 & Monitor its operation, consider over-reliance, override or reverse the output and intervene on the operation or interupt the system (e.g., through a “stop” button). \\
    Cybersecurity & E08 & Cybersecurity and resilience to prevent and control for attacks. \\
    Register & E09 & Register their systems in an EU-wide database in any external AI database (e.g., EU-wide database). \\
    Compliance & E10 & Conformity assessment and management process. \\
    \hline
    \end{tabular}
    \caption{10 examples requirements identified from the EU AI Act with a short summary; the details can be found in \textit{the EU AI Act document} \cite{euaiact}.}
    \label{tab:high-risk requirement}
    \normalfont
\end{table}

\textbf{2) Map the requirements with risk questions in the RAI Question Bank.}
By mapping these requirements to specific questions in the RAI Question Bank, we aim to provide a comprehensive framework for assessing compliance and ensuring that AI systems are developed and used responsibly.
This step involves a detailed analysis of each requirement from the Act and its alignment with the ethical principles and risk factors outlined in the RAI Question Bank.
Accordingly, each requirement was carefully interpreted to understand its implications and scope. We then matched each requirement with one or more relevant questions from the RAI Question Bank that address similar concerns or objectives. This mapping process ensures that for every legal requirement, there is a corresponding risk question that can be used to evaluate whether an AI system meets the necessary requirements.
For example, a requirement related to \textit{technical specification} might be mapped to questions about the AI system's transparency and explainability, the documentation of its decision-making processes, and the accessibility of its outputs to users. Similarly, a requirement focused on data management would be mapped to questions about data privacy, security measures, and data management practices.

This systematic approach ensures that all legal requirements are comprehensively covered by the selected questions for risk assessment. It also facilitates a clear and structured method for practitioners to evaluate and ensure their AI systems comply with the EU AI Act, promoting ethical and responsible AI development.

\begin{comment} blocked due to page limit
For this mapping, we designed a simple schema as shown in Figure \ref{fig:EU mapping}.
In the figure, we also show example screenshots for each entity.

\begin{figure*}[htb]
    \centering
    \includegraphics[width=0.9\textwidth]{Fig/EUMapping.pdf}
    \caption{The designed schema for mapping the requirements to the RAI Question Bank questions.}
    \label{fig:EU mapping}
\end{figure*}

\begin{itemize}
    \item \textit{EU-base} is a core entity that includes the identifier (ID), requirement description and corresponding section in the document.

    \item \textit{EU-QB} represents a mapping table that comprises the requirement ID and question ID. Each requirement has at least one or multiple mappings in this table.

    \item \textit{EU-QB-sub} includes two question IDs to demonstrate subsequent questions. Each question in \textit{EU-QB} can have zero or one mapping.
\end{itemize}

This schema is straightforward yet facilitates effective navigation of risk questions that cover all compliance requirements.
\end{comment}

\textbf{3) Score the AI system.}
Users can provide answers to the questions using a simple binary metric based on their situation. 
\begin{itemize}
  \item \textit{Yes}: Indicates full compliance or agreement.
  \item \textit{No}: Indicates non-compliance or disagreement.
  \item \textit{N/A (Not Applicable)}: Indicates that the question does not apply to the specific case or context.
\end{itemize}

For the question, \textit{Do you establish an AI risk management system to conduct ongoing risk assessment and treatment?}, users can answer \textit{Yes} if they have implemented a structured system for ongoing assessment and treatment of AI risks. Answering \textit{No} indicates the absence of such a system. 
\textit{N/A} should be selected if the organization does not engage in AI activities or if this specific practice is not applicable to their operations. 
Scores are assigned based on whether the answer reflects compliance (Yes) or non-compliance (No), with N/A indicating exemption from scoring.
This approach provides clarity on how to interpret and score responses in a compliance assessment using a simple binary metric. 
Adjustments can be made based on specific scoring criteria or compliance frameworks relevant to the user context.

The following shows how to calculate a simple score using the 21 questions.

\( Q_i \) represents the score for each of the 21 risk questions and \( S \) denotes the final compliance score of the high-risk AI, against the EU AI Act.
We assigned an equal weight of 1 for each question. Thus, the formula with weights can be expressed as follows (\ref{eq:compliance score 1}).

\begin{equation}
S = \sum_{i=1}^{10} w_q Q_i    
\label{eq:compliance score 1}
\end{equation}

where \( w_q = 1 \) is the weight for each question.

The compliance level \( L \) is determined as follows (\ref{eq:compliance score 2}).

\begin{equation}
L = 
\begin{cases} 
\text{Full Compliance} & \text{if } S = 21 \\
\text{Partial Compliance} & \text{if } T \leq S < 21 \\
\text{Non-Compliant} & \text{if } 0 \leq S < T \\
\end{cases}    
\label{eq:compliance score 2}
\end{equation}

In this formula, \( T \) represents the threshold score.
Scores from 0 to \( T \) - 1 indicate non-compliance.
Scores from \( T \) to 20 indicate partial compliance.
A score of 21 indicates full compliance.

Users can adjust the value of \( T \) based on their specific compliance criteria and context.

\subsection{Empowering Trust of AI Agents} 

%This section addresses another potential and promising area, \textit{AI agents} where the RAI Question Bank can play a focal role in the AI space.

%As introduced earlier, 
AI agents can autonomously complete user tasks by interacting with users and other parties, such as external foundation models, tools, and other agents \cite{castelfranchi1998modelling}. 
These agents share more control with humans (e.g., the agent owner, users) and can operate autonomously without direct human intervention (e.g., use of a credit card to make purchases on behalf of the user).
This underscores the need for responsible development and use of AI agents, with ongoing risk assessments to prevent serious incidents and harms to individuals, society, and the environment \cite{lu2023towards}. %Additionally, when users set specific goals for the agent, the underlying foundation model is queried to achieve these goals. 
It is crucial to prevent the foundation model from being influenced by adversarial inputs and ensure it does not generate harmful or undesirable outputs for users or other components \cite{liu2024agent}.
To foster trust in AI agents, both the agents themselves and the foundation models they interact with must be trustworthy.
The RAI Question Bank can be applied to the both sides.
\begin{comment} blocked due to page limit
Figure \ref{fig:agent} shows the conceptual diagram for this application.

\begin{figure*} [htb]
    \centering
    \includegraphics[width=\textwidth]{Fig/Agent.pdf}
    \caption{The potential application of the RAI Question Bank for foundation model-based AI agents.}
    \label{fig:agent}
\end{figure*}
\end{comment}

\textbf{Risk assessment of AI agents.}
We have adopted \textit{RAI plugins} from the AI agent design patterns proposed in 2023 \cite{lu2023towards}.
\textit{RAI plugins} comprises five agent components: Continuous
risk assessor, Black box recorder, Explainer, Multimodal guardrails, and AIBOM registry. % Co-versioning registry has been removed as it is about foundation model.
These components serve as an RAI role to assess and monitor risks and to record all relevant data and information of the AI agent.
We have interpreted the roles and requirements of these components and identified relevant risk questions from the RAI Question Bank as follows (Table \ref{tab:agent}).

\begin{table} [htb]
    \centering
    \footnotesize
    \begin{tabular}{p{0.15\textwidth}p{0.25\textwidth}p{0.45\textwidth}p{0.05\textwidth}}
    \hline
    RAI component & Role/requirement & Risk question & Source\\ 
    \hline
    \multirow{1}{0.15\textwidth}{Continuous risk assessor} & \multirow{1}{0.25\textwidth}{Monitor and assess AI risks} & 1.Does the agent have risk management mechanisms to conduct ongoing risk assessment and treatment? & ISO \\
     & & 2.Does the agent conduct AI risk assessment? & ISO\\
     & & 3.Does the agent implement the AI risk treatment plan? & ISO\\
    \multirow{1}{0.15\textwidth}{Black box recorder} & \multirow{1}{0.25\textwidth}{Record and share the runtime data} & 4.Does the agent have a logging function to record interactions or data? & EU\\
    & & 5. Does the agent record all the recommendations or decisions made by the system? & AIA \\
     & & 6.Does the agent share the records with stakeholders when required? & AIA \\
   \multirow{1}{0.15\textwidth}{Explainer} & \multirow{1}{0.25\textwidth}{Articulate the agent’s roles, capabilities, limitations, the rationale behind its intermediate or final outputs.} & 7. Can the agent produce outcomes that all users can understand? & EU \\
   && 8. Is the agent designed with interpretability in mind from the start? & EU \\
   && 9. Do you document and provide technical specification including the intended purposes, potentially beneficial uses, limitations and outputs of the agent? & NIST \\
   \multirow{1}{0.15\textwidth}{Multimodal guardrail} & \multirow{1}{0.25\textwidth}{Control the inputs and outputs of foundation models to meet specific requirements.} & 10. Is the agent designed to have an appropriate level of oversight and control for foundation models?  & EU \\
   & & 11. Does the agent monitor and control adversarial inputs, harmful or undesirable outputs to users and other components? & EU \\
   \multirow{1}{0.15\textwidth}{AIBOM registry} & \multirow{1}{0.25\textwidth}{Records their supply chain details including AI risk metrics or verifiable RAI credentials. } & 12. Do you register all components including external tools, agents and models? &  - \\
   & & 13. Does the AI agent verify the provenance of the components and manage their execution? & - \\
    \hline
    \end{tabular}
    \caption{AI agent RAI plugins and the RAI Question Bank questions.}
    \label{tab:agent}
    \normalfont
\end{table}

The \textit{Continuous Risk Assessor} is responsible for the ongoing monitoring and management of AI risks. It requires capabilities to perform AI risk assessments and implement risk treatments to address identified risks. This component is primarily associated with \textit{Accountability} principle, which focuses on comprehensive responsibility of AI management.
We suggest three key questions for this component which are based on the ISO/IEC AI management standard.

The \textit{Black Box Recorder} is an essential component for ensuring the traceability of AI systems \cite{mora2021traceability}. It serves the crucial function of recording and storing comprehensive information about the AI system’s operations, decisions, and performance. This includes logging inputs, outputs, processing steps, and any anomalies encountered during the system's operation. By maintaining a detailed record, it supports accountability and facilitates audits, making it possible to review and understand the system’s behavior retrospectively. 
More importantly, this component enhances transparency for stakeholders, enabling them to verify that the AI system operates as intended and adheres to relevant regulatory and ethical standards \cite{khan2022impact}. Additionally, it helps in assigning clear responsibility by documenting who interacted with the system and how, thereby contributing to a robust governance framework \cite{kroll2021outlining}.
There are three questions for this component adapted from EU Assessment List for Trustworthy Artificial Intelligence and Canada Algorithmic Impact Assessment framework.

\textit{Explainer} plays a crucial role in providing important information about the AI agent, including its roles, capabilities, limitations, and outputs. This component is specifically designed to enhance transparency and explainability practices. Key considerations for this component include that AI systems should produce understandable outcomes, that explainability should be integrated from the very early design stages, and that all necessary information should be thoroughly documented \cite{markus2021role}. By addressing these points, the \textit{Explainer} ensures that stakeholders have a clear understanding of how the AI system operates and can trust its decisions.

\textit{Multimodal guardrail} is designed to prevent inappropriate multimodal inputs from being sent to the foundation model and to monitor and control the outputs from the model \cite{sharma1998toward}. It should be capable of processing various forms of data, such as text, audio, and video, to support comprehensive control \cite{liu2024agent}. Positioned in the middle layer between the foundation model and other components, this component serves a crucial role in the operational loop.

\textit{AIBOM Registry} refers to the storage for all necessary information associated with the AI supply chain, including procurement information \cite{bi2024way}. 
The AIBOM registry can include not only internal components and their information but also external tools, agents, and models \cite{stalnaker2024boms}. 
To ensure RAI practices, the AI agent needs to verify the provenance of the components. Accordingly, this component is strongly interrelated with the \textit{Black Box Recorder}.

\textbf{Risk assessment of foundation models.}
As a key external entity, a foundation model generates outputs in response to the AI agent's prompts. 
To mitigate risks and minimize negative impacts on AI agents, model providers should conduct ongoing risk assessments and ensure compliance with regulations and standards \cite{bommasani2021opportunities}. 
From the AI agent's perspective, it is essential to have confidence that the foundation model is developed and deployed responsibly.
To support this, we have identified a set of essential requirements for verifying foundation models, primarily derived from the EU AI Act. %Additionally, we suggest relevant questions based on the RAI Question Bank to check these requirements.
Table \ref{tab:foundation model} shows the requirement list and corresponding questions to address the requirements from the AI agent's perspective.

\begin{table}[htb]
    \centering
    \footnotesize
    \begin{tabular}{p{0.15\textwidth}p{0.2\textwidth}p{0.4\textwidth}p{0.15\textwidth}}
    \hline
    Category & Requirement & Risk question & Disclosure \\ 
    \hline
    \multirow{1}{0.15\textwidth}{Risk management} & \multirow{1}{0.2\textwidth}{Risk management and involve external experts} & 1.Does the model provider conduct ongoing risk assessment and treatment? & \multirow{1}{0.15\textwidth}{Risk assessment and mitigation report} \\
     & & 2.Does the model provider involve external experts in risk management? & \\
    \multirow{1}{0.15\textwidth}{Data governance} & \multirow{1}{0.2\textwidth}{Suitability of the data sources, possible biases and appropriate mitigation} & 3.Does the model provider assess the quality of the data sources used for training? & \multirow{1}{0.15\textwidth}{Summaries of training data used} \\
    & & 4. Does the model provider document detailed summary of the use of training data protected under copyright law? & \\
    Documentation & \multirow{1}{0.2\textwidth}{Documentation for downstream providers} & 5. Does the model provider document and provide technical specification? & \multirow{1}{0.15\textwidth}{Technical specification} \\
    & & 6. Does the model provider provide intelligible instructions for use? & User instructions \\
    \multirow{1}{0.15\textwidth}{Environmental impact} & \multirow{1}{0.2\textwidth}{Measurement of environmental impact} & 7. Does the model provider assess and document environmental impact and sustainability of AI model training and management activities? & \multirow{1}{0.15\textwidth}{Energy usage, carbon emission, tones of waste} \\
    \multirow{1}{0.15\textwidth}{Quality management} & Model evaluation with documented analysis and extensive testing & 8. Does the model provider conduct and document model evaluation through its lifecycle? & Testing report \\
    \hline
    \end{tabular}
    \caption{Risk assessment for foundation models from the AI agent developer perspective.}
    \label{tab:foundation model}
    \normalfont
\end{table}

It is crucial that model providers conduct ongoing risk assessments and implement effective treatment strategies. 
To ensure a robust risk management process, involving external experts is recommended. 
The evidence supporting compliance with these requirements can be documented in a comprehensive \textit{risk assessment and mitigation report.}

Effective data governance necessitates that model providers rigorously evaluate the quality of training data sources to ensure they are suitable and free from bias. It is also critical to maintain thorough documentation of any training data protected by copyright law. 
%Key questions to consider can include \textit{Does the model provider assess the quality of the data sources used for training?} and \textit{Does the model provider document detailed summaries of the use of training data protected under copyright law?}
Evidence of adherence to these standards is found in detailed \textit{summaries of the training data employed.} 
As known solutions to support this, \textit{Datasheets for datasets} \cite{gebru2021datasheets} and \textit{Data bill of materials} can be considered.

%\yue{Data Bill of Materials? Or data sheets/card as they are more mature solutions..}

In the area of documentation for downstream providers, it is important for model providers to produce and disseminate comprehensive technical specifications along with clear, understandable instructions for use. Providing detailed \textit{technical documentation and user instructions} is essential for the proper implementation and utilization of the models. Model providers should transparently demonstrate these practices through the \textit{technical specifications and user manuals} made available to users (i.e., foundation model-based agent developers).

The foundation model providers should assess and document \textit{the sustainability of AI model} training and management activities. 
This includes measuring factors such as energy usage, carbon emissions, and tons of waste generated \cite{martiny2023towards}. 
Model providers should also consider additional metrics related to model/data size, complexity, training time, and computational performance (e.g., FLOPs), as these factors can indirectly impact the environment. 
Providing detailed assessments and documentation of these metrics ensures transparency and accountability in their environmental practices.

\textit{Quality management} mandates rigorous evaluations and meticulous documentation throughout the lifecycle of AI models \cite{euaiact}. 
This process entails thorough testing and analysis to ensure reliability, accuracy, and adherence to predefined performance criteria. 
The testing report serves as crucial evidence, detailing the evaluation methods used, findings, and any adjustments made over the lifecycle. 
These practices enable the model providers to enhance transparency and facilitate continuous improvement in their model development processes, thereby allowing agent developers to ascertain the model's trustworthiness.

\begin{comment} blocked due to page limit
The main purpose of this section is to demonstrate how our RAI Question Bank can be used for the risk assessment of AI agent systems, rather than developing a comprehensive risk assessment framework. This section, therefore, may lack detailed methodologies or extensive examples, as its focus is on illustrating the practical application of the question bank.
\end{comment}

\begin{comment}
\subsection{Smart Risk Assessment Tool}
\sunny{The following is from the previous QB paper. There is limited information/description. Need to include more or add some screenshots if can. ---> will contact Mulong later and consider this section for journal submission.}

The Smart Risk Assessment Tool (SRA) is an innovative chatbot designed to provide users with answers to their questions regarding AI risk assessment. 
The development of SRA draws upon the analysis of several hundreds of AI incident cases, which have been collected and analyzed to construct a comprehensive knowledge graph.
The current version of SRA serves as a prototype, demonstrating its capabilities and potential as a proof of concept. 
It utilizes ChatGPT as a foundation model to generate responses and future iterations will involve the utilization of fine-tuned local models, enhancing the accuracy and contextual relevance of the tool.
\end{comment}

%\subsection{Threats to validity}
%will write later for journal submission

\section{Conclusion} \label{sec:conclusion}

In this study, we presented the RAI Question Bank, developed through comprehensive approaches, including a systematic literature review, an in-depth survey of existing RAI frameworks, and multiple case studies.
The RAI Question Bank represents a significant step forward in the structured risk assessment of AI systems. Its hierarchical structure, organized into different levels (Level 1-3), ensures a comprehensive evaluation by linking high-level principles with detailed, technical questions. By organizing RAI themes and risk questions around eight AI ethics principles, it helps effectively identify potential risks, ensure compliance with emerging regulations, and enhance overall AI governance.

Throughout this paper, we have detailed the methodology used to develop and refine the RAI Question Bank, including the incorporation of feedback from case studies and iterative improvements. The examination of the inclusion of specialized instructions and metrics addressed the need for evidence-based quality assessments of questions and showed its potential for mitigating the risk of a checkbox mentality.

While the current version of the RAI Question Bank is comprehensive, we have a plan for further enhancement, particularly in integrating more assessment instructions and quality metrics to support evaluators. 
The next steps will focus on expanding the depth and breadth of the question bank, ensuring that it remains a robust tool for fostering trustworthy and responsible AI development.

In conclusion, the RAI Question Bank offers valuable insights and serves as a knowledge base, functioning as a core resource and tool for organizations, while promoting a thorough and structured approach to risk assessment. It supports the responsible use and development of AI systems that are not only innovative but also aligned with ethical principles and societal values.

%Bibliography
\bibliographystyle{unsrt}  
\bibliography{main}

\end{document}